\documentclass{article}

\usepackage{arxiv}

\usepackage[utf8]{inputenc} 
\usepackage{enumerate}
\usepackage[T1]{fontenc}    
\usepackage{hyperref}       
\usepackage{url}            
\usepackage{booktabs}       
\usepackage{amsfonts}       
\usepackage{nicefrac}       
\usepackage{microtype}      
\usepackage{lipsum}		
\usepackage{graphicx}
\usepackage{natbib}
\usepackage{doi}
\usepackage{amsmath}
\usepackage{bbm}
\usepackage[usenames,dvipsnames,svgnames,table]{xcolor}
\usepackage{hyperref}
\hypersetup{
     colorlinks   = true,
     citecolor    = teal
}
\usepackage{multirow}
\usepackage{multicol}
\usepackage{makecell}
\usepackage{threeparttable}
\usepackage{adjustbox}
\usepackage{booktabs}
\usepackage{caption}
\usepackage{url}

\title{Bayesian Joint Modeling of Longitudinal Symptomatology Scale Responses and Fall Outcomes via Heterogeneous Latent Transition Analysis}

\author{ \href{https://orcid.org/0009-0009-6361-9760}{\includegraphics[scale=0.06]{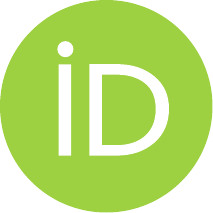}\hspace{1mm}Mingyan~Yu} \\
	Department of Biostatistics\\
	University of Michigan\\
	Ann Arbor, MI 48109 \\
	\texttt{myanyu@umich.edu} \\
	\And
	\href{https://orcid.org/0000-0001-7582-669X}{\includegraphics[scale=0.06]{orcid.pdf}\hspace{1mm}Zhenke~Wu} \\
	Department of Biostatistics\\
	University of Michigan\\
	Ann Arbor, MI 48109 \\
	\texttt{zhenkewu@umich.edu} \\
    \And Michelle M.~Hood \\
	Department of Epidemiology\\
	University of Michigan\\
	Ann Arbor, MI 48109 \\
	\texttt{mmhood@umich.edu} \\
    \And Carrie A. Karvonen-Gutierrez \\
	Department of Epidemiology\\
	University of Michigan\\
	Ann Arbor, MI 48109 \\
	\texttt{ckarvone@umich.edu} \\
    \And Fan Bu \\
	Department of Biostatistics\\
	University of Michigan\\
	Ann Arbor, MI 48109 \\
	\texttt{fbu@umich.edu} \\
    \And 
    \href{https://orcid.org/0000-0003-4128-5527}{\includegraphics[scale=0.06]{orcid.pdf}\hspace{1mm}Michael R.~Elliott} \\
	Department of Biostatistics\\
	University of Michigan\\
	Ann Arbor, MI 48109 \\
	\texttt{mrelliot@umich.edu} \\
}

\renewcommand{\shorttitle}{Joint Modeling of Longitudinal Symptomatology Scale Responses and Fall Outcomes}

\hypersetup{
pdftitle={Joint Modeling of Longitudinal Symptomatology Scale Responses and Fall Outcomes},
pdfsubject={Joint Model},
pdfauthor={Mingyan~Yu, Zhenke~Wu, Michelle M.~Hood, Carrie Karvonen-Gutierrez, Fan~Bu, Michael R.~Elliott},
pdfkeywords={Bayesian hierarchical model, Joint multivariate categorical modeling, Latent transition analysis, Midlife falls, Women's Health.},
}

\begin{document}
\maketitle

\begin{abstract}
	The Study of Women’s Health Across the Nation (SWAN) has followed women for over 30 years, from midlife premenopause until later life. The study has 16 surveys at approximately 2 years intervals that cover a wide range of physical and psychological symptoms. These multivariate categorical survey responses potentially contain rich health-related information. Temporal trajectories of the survey responses can be characterized by both the responses profiles and the evolving dynamics of the responses over time. To capture those two features and investigate how they inform subsequent health outcomes, we propose a joint multi-layer latent transition model. We combine a latent transition model that classifies individuals based on their response profiles over time with an additional layer of clustering of these latent class transition sequences, with the goal of connecting these cluster profiles with health outcomes: in this application, self-reported falls. In addition, we evaluate the operating characteristics of the method through simulation studies.
\end{abstract}


\section{Introduction}
\label{sec:intro}
Falls are a well-recognized public health concern, especially among older adults \citep{o1993incidence, gill2005population}. Evidence suggests that females are more likely than males to experience falls and fall-related injuries across all age groups \citep{ylitalo2016body}, highlighting the importance of preventive interventions for females. Although most research has focused on females aged 65 years and older, emerging studies suggest a rising prevalence of falls during midlife. For example, \citet{peeters2018should} reported that the prevalence of falls increased from 8.7\% among females aged 40-44 to 29.9\% among those aged 60-64 years across four population-based cohort study. Moreover, \citet{karvonen2020midlife} found that midlife females who experienced recurrent falls had more than a three-fold higher hazard of morality within 10 years compared to others. Together, these findings emphasize the necessity to initiate early preventive efforts at middle ages.

A wide range of risk factors for falls among midlife females has been identified \citep{white2018fall, peeters2019comprehensive}, including obesity, poor physical functioning, stiff/painful joints, depression, osteoporosis, severe tiredness, and other chronic conditions. These factors often present in clusters, suggesting that falls arise from a dynamic integration of multiple risk factors rather than an isolated cause. Consequently, examining risk factors jointly as clusters may better capture the underlying risk profiles. However, most existing studies rely on cross-sectional measurements and do not account for how these risk factors change over time. Understanding their longitudinal trajectories may provide additional insights into midlife falls and support earlier intervention. To address this gap, we analyze the longitudinal symptomatology data from the Study of Women's Health Across the Nation (SWAN) \citep{swan}, which contains repeated categorical measures of physical, social and psychological symptoms along with self-reported fall outcomes. Our study has two primary aims: (1) to identify potential symptom clustering and characterize their longitudinal dynamics while accounting for population heterogeneity; and (2) to examine how these symptom profiles and their temporal dynamics are associated with the occurrence of falls, particularly recurrent falls.

Conceptually, symptom clustering can be approached from either a variable-centered or a person-centered view \citep{howard2018variable}. Variable-centered approaches, such as factor analysis \citep{huang2016symptom, seppala2020factor}, group interrelated symptoms according to similarity measures under a homogeneous population assumption \citep{kim2008statistical}. In contrast, person-centered approaches identify subgroups of individuals with similar patterns of symptoms within a heterogeneous population. Since our goal is to characterize individual symptom profiles and their dynamics over time, we adopt a person-centered approach and develop an extension of latent transition analysis (LTA) that we term heterogeneous latent transition analysis or HTLA.

Standard LTA \citep{collins2013latent} is a longitudinal extension of latent class analysis (LCA), a finite mixture model that groups individuals based on their response patterns. By incorporating repeated symptom measurements, LTA models both latent class memberships and the transitions between classes over time through Markovian transition probability matrices \citep{collins2013latent}. LTA has been widely applied in psychological, social and health studies. For example, \citet{harlow2017not} identified six symptoms classes among midlife women based on physical, menopausal and psychological symptoms and found substantial stability in class memberships over time. Similarly, \citet{ni2017changes} identified three depression subgroups from the China Health and Retirement Longitudinal Study and observed considerable stability, with a moderate fraction transitioning from ``mild" and ``severe" depression classes to the ``lack of positive affect" class.

Although LTA has been widely used to identify latent subgroups and characterize their transition dynamics, relatively few studies have examined whether these transition patterns are predictive of subsequent health outcomes. For example, based on two waves of the Chinese Longitudinal Healthy Longevity Survey, \citet{wang2024latent} identified three latent classes of social participation and found that transitions to classes with higher levels of social participation were associated with reduced depression. However, such approaches become impractical when the number of possible transition patterns grows with the number of visits and latent classes. An alternative approach introduces a mover-stayer second-order latent variable \citep{goodman1961statistical} within the LTA framework, which partitions individuals into subgroups who either remain in the same class over time (``stayers") or transition between classes (``movers"). \citet{zammit2020latent} applied this approach to neuropsychological measures from the Rush Memory and Aging Project and found that “movers” had a substantially higher risk of developing Alzheimer’s disease. Though appealing, this approach imposes a restrictive assumption that a subgroup of individuals never change class and does not account for the heterogeneity in transition patterns among ``movers". 

To address these limitations, we introduce an additional layer of heterogeneity by clustering the latent class sequences obtained from LTA. Our approach is motivated by the model-based pattern clustering for categorical time series \citep{fruhwirth2010model, mcnicholas2016model}, where the observed series are modeled using a finite mixture of first-order Markov processes; this approach has been used to model, e.g., wage morbidity \citep{fruhwirth2010model} and daily activity patterns \citep{zhou2021you}. In contrast, the sequences in our setting arise from an underlying latent processes and are unobserved. By incorporating both latent classes and sequence clusters, our model captures heterogeneity at multiple levels. 

Building on this framework, we further examine how latent states and transition dynamics relate to subsequent outcomes by linking the outcome to sequence cluster membership and latent states at certain time points through regression models. A key technical challenge is properly accounting for the uncertainty in both latent classes and sequence clusters to avoid biased inference. In the simpler setting of relating only the latent class membership to a distal outcome, several approaches has been proposed. The naive classify-analyze approach \citep{clogg1993latent} uses modal class assignments as predictors but ignores the classification error, thus introducing measurement error bias. To address this issue, several correction methods have been developed, including the pseudo-class draw approach \citep{bandeen1997latent, wang2005residual}, the maximum-likelihood based three-step approach \citep{vermunt2010latent}, the Bolck-Croon-Hagenaars (BCH) approach \citep{bolck2004estimating}, and a model-based approach that leverages Bayes’ theorem \citep{lanza2013latent}. 
However, the multi-layer and longitudinal structure of our method introduces additional complexity for bias correction. We therefore adopt a Bayesian joint modeling framework that fully propagates uncertainty from both latent classes and sequence clusters into the outcome model. Joint modeling has been widely used to analyze longitudinal measurements and cross-sectional outcomes for its flexibility and ability to yield unbiased association estimation. For example, 
\citet{elliott2020methods} examined acculturation behaviors using LCA with the latent class membership associated with depression outcome, demonstrating the advantages of joint modeling in uncertainty quantification. 

This paper has three major innovations: 1) extending the LTA model to allow for heterogeneity (HLTA); 2) developing a Bayesian joint model to account for measurement error in relating the HTLA sequence classes to the outcome, and 3) developing methods to account for label switching at both the latent class and latent transition layers in the MCMC algorithm. The manuscript is organized as follows: Section \ref{sec:method} develops the statistical details of the proposed joint modeling approach for longitudinal symptomatology data and subsequent cross-sectional outcomes in a Bayesian framework; Section \ref{sec:simulation} evaluates the performance of the method through simulation studies; Section \ref{sec:application} applies our proposed method to the longitudinal item-responses and fall outcomes from SWAN to address the motivating scientific question, and Section \ref{sec:discussion} concludes with a summary and discusses potential limitations and directions for future work.

\section{Methods}
\label{sec:method}

\subsection{Notation}\label{method:notation}

Let $\boldsymbol{X}_i$ be a $T\times K$ matrix of categorical item-responses for individual $i$ collected over $T$ visits, with $K$ representing the total number of items administered. The $t$-th row of $\boldsymbol{X}_i$, $\boldsymbol{X}_{it}=(X_{it1},\ldots,X_{itK})$, contains the responses of individual $i$ at visit $t$, for $t=1,\ldots,T$. We allow individuals to miss some visits, therefore their item-responses can be all missing at those visits. We also assume that not all items are necessarily administered at every visit, thus some item-responses may be missing at some visits by design. For the $k$-th item, its responses can take values in $\{1,\ldots,d_k\}$, $d_k\ge 2$. Let scalar $Y_i$ denote the observed cross-sectional outcome for individual $i$, and let vector $\boldsymbol{Z}_i$ denote the baseline covariates.  

\subsection{Proposed Joint Modeling Framework}\label{method:joint}

Our proposed joint modeling framework consists of two connected components. First, we propose a sequence clustering method based on the latent transition analysis to capture the individual-level longitudinal patterns of item-responses collected over $T$ visits. Then we are interested in associating the cross-sectional outcome $Y_i$ with the sequence clusters and item-response profiles to draw inferences on how the longitudinal item-response trajectories and profiles inform the outcome.

\subsubsection{Longitudinal Patterns of Item-Responses}\label{joint:longitudinal}

We begin by summarizing individuals' multivariate item-responses at a single visit using latent class analysis (LCA). LCA defines classes through class-specific item-response probabilities and assumes conditional independence of item-responses given the latent class. Let $L<\infty$ denote the number of latent classes and let $\eta_i\in \{1,\ldots,L\}$ represent the latent class membership for individual $i$. For the $k$-th item with $d_k$ categories, we denote the class-specific item-response probabilities as $p_{lj}^{(k)}=\text{Pr}(x_{itk}=j|\eta_i=l), j=1,\ldots,d_k, l=1,\ldots,L$, which are shared by individuals within the same latent class. The LCA framework can be extended to longitudinal multivariate item-responses collected over $T$ visits, yielding a sequence of latent classes $\{\eta_{i,1},\ldots,\eta_{i,T}\}$, where $\eta_{i,t}$ denotes the latent class membership of individual $i$ at visit $t$. We assume that the number of latent classes and their item-response profiles remain constant over time. Individuals are allowed to transition between latent classes across visits, with the temporal dependency modeled through an $L\times L$ transition matrix $\boldsymbol{\tau}$, where $\tau_{l,l^\prime}=\text{Pr}(\eta_{i,t+1}=l^\prime|\eta_{i,t}=l)$ represents the probability of transitioning from class $l$ to class $l^\prime$, with $\sum_{l^\prime=1}^L\tau_{l,l^\prime}=1$. We further assume the transition matrix is time-invariant and shared across individuals, although extensions to time-varying transition matrices are possible. Let $\boldsymbol{\pi}=(\pi_1,\ldots,\pi_L)$ denote the vector of initial class probabilities at visit 1, with $\pi_l=\text{Pr}(\eta_{i,1}=l)$ and $\sum_{l=1}^L\pi_l=1$. Together, the longitudinal item-response trajectories are characterized by $\boldsymbol{\pi}$ and $\boldsymbol{\tau}$.

\subsubsection{Extending LTA To Allow for Heterogeneous Transition Patterns}\label{joint:extending}

While LTA captures individual-level item-response dynamics, it does not explicitly account for heterogeneity in the transition patterns across individuals. To address this limitation, we introduce an additional layer of clustering on the latent class sequences. Specifically, we assume that individuals can be grouped into $R<\infty$ sequence clusters, with sequence cluster membership denoted by $\lambda_i\in \{1,\ldots,R\}$ and $\text{Pr}(\lambda_i=r)=b_r$ with $\sum_{r=1}^R b_r=1$. Within each sequence cluster $r$, individuals share common initial class probabilities $\boldsymbol{\pi}^{(r)}$ and transition matrix $\boldsymbol{\tau}^{(r)}$. The item-response probabilities $p_{lj}^{(k)}$ are assumed to be invariant across time and sequence clusters, ensuring a consistent interpretation of the latent classes. Built upon LTA and latent class sequence clustering, the longitudinal multivariate item-response model is specified as
\begin{equation}
\begin{gathered}
    x_{itk}|\eta_{i,t}=l\sim \text{Multinomial}(p_{l1}^{(k)},\ldots,p_{ld_k}^{(k)}), k=1,\ldots,K, l=1,\ldots,L; \\
    \lambda_i\sim \text{Multinomial}(b_1,\ldots,b_R), \text{independently for } i=1,\ldots,N; \\
    \eta_{i,1}|\lambda_i=r\sim \text{Multinomial}(\pi_1^{(r)},\ldots,\pi_L^{(r)}), \\
    \eta_{i,t+1}|\eta_{i,t}=l,\lambda_i=r\sim \text{Multinomial}(\tau_{l,1}^{(r)},\ldots,\tau_{l,L}^{(r)}), r=1,\ldots,R, t=1,\ldots,T-1, l=1,\ldots,L.
\end{gathered}
\end{equation}

\paragraph{Prior Specification}\label{joint:prior}

We assign Dirichlet priors to all the multinomial probability vectors in the model for their conjugacy and flexibility in modeling probabilities: 
\begin{equation}
    \begin{gathered}
        b_1,\ldots,b_R\sim \text{Dirichlet}(\zeta_1,\ldots,\zeta_R); \\
        \pi_1^{(r)},\ldots,\pi_L^{(r)}\sim \text{Dirichlet}(a^{(r)}_1,\ldots,a^{(r)}_L), \\
        \tau_{l,1}^{(r)},\ldots,\tau_{l,L}^{(r)}\sim \text{Dirichlet}(\psi_{l1}^{(r)},\ldots, \psi_{lL}^{(r)}), r = 1,\ldots,R, \\
        p_{l1}^{(k)},\ldots,p_{ld_k}^{(k)}\sim \text{Dirichlet}(\omega^{(k)}_{l1},\ldots,\omega^{(k)}_{ld_k}), k=1,\ldots,K, l=1,\ldots,L.
    \end{gathered}
\end{equation}

In our simulation and application implementation, we choose to use weakly informative priors by setting all hyper-parameters to 1. More informative prior can be used if prior knowledge regarding the expected sequence cluster and cluster-specific latent class distributions or item-response probabilities is available. 

\subsection{Cross-Sectional Outcomes}\label{method:outcome}

We model the association between the longitudinal multivariate item-responses and the cross-section outcome $Y_i$ through the sequence cluster $\lambda_i$ and the latent class at the final visit $\eta_{i,T}$. The inclusion of $\lambda_i$ as a predictor aims to examine how longitudinal item-response trajectories impact the outcome, while $\eta_{i,T}$ captures the effect of the most recent item-response profile. The outcome model is given by 
\begin{equation}
    \begin{gathered}
        g(E[Y_i])=\gamma(\boldsymbol{W}_i, \boldsymbol{\beta}), \ \ \boldsymbol{W}_i=(1,\lambda_i, \eta_{i,T}, \boldsymbol{Z}_i^\top)^\top.
    \end{gathered}\label{eq:4.2}
\end{equation}
where $g(\cdot)$ represents a link function. The vector $\boldsymbol{\beta}$ represents the regression coefficients, for which we assume independent weakly informative $\mathcal{N}(0,\sigma^2)$ priors with $\sigma=10$. Flexible specification for the functional form $\gamma(\cdot)$ of the predictors $\boldsymbol{W}_i$ can be assumed, and in our simulation studies and application, we adopt a linear combination of the predictors. 

\subsection{Posterior Inference}\label{method:inference}

Our method is implemented using the \textsf{NIMBLE} \citep{nimble-article:2017, nimble-software:2025} package in \textsf{R}. \textsf{NIMBLE} has the nice feature to compile a BUGS model using \textsf{C++}, which substantially speed up the sampling process. For the simulation studies, we run three chains each with 10,000 iterations and 5,000 burnin. For the application study, we run one chain with 24,000 iterations and 12,000 burnin. Convergence is assessed by visual inspection of the traceplots and the Gelman-Rubin $\hat{R}$ measure \citep{gelman2013bayesian}. For label switching issue commonly seen in mixture models, we adopt and extend Stephens' relabeling algorithm \citep{stephens2000dealing} in an efficient way to relabel the posterior samples. The standard Stephens' relabeling algorithm for latent classes involves considering the posterior classification probability all of the possible label assignment permutations for each MCMC draw, and maximizing this posterior classification probability across all draws iteratively. Directly obtaining this posterior classification probability at the sequence cluster required marginalizing across the latent class assignments at each time point, which quickly becomes intractable as the number of time points grows beyond 3 ot 4. Thus this extension separates the ``untangling'' of the latent class assignment and sequence cluster assignments via a forward-backwards algorithm (with respect to the label permutations).  The detailed algorithm can be found in Supplementary Materials Section 1.

\section{Simulation}\label{sec:simulation}

Before applying the proposed joint modeling approach to the longitudinal scales dataset from SWAN, we conduct simulation studies to evaluate the performance of the proposed method and identify an appropriate model selection criterion.  

\subsection{Data Generation}\label{sim:gen}

We start with generating the longitudinal item-responses data through three steps. We consider $R=3$ underlying sequence clusters and $L=4$ underlying item-response latent classes. The observed data consist of responses to $K=30$ items collected over $T=10$ visits for $N=1500$ individuals. First, for each individual $i$, we sample the sequence cluster membership $\lambda_i$ from a $\text{Multinomial}(0.20, 0.35, 0.45)$ distribution. Given $\lambda_i$, the latent class sequence $\{\eta_{i,1},\ldots,\eta_{i,T}\}$ is generated from a first-order Markov process. The initial latent class at visit 1 in sampled as:
\begin{equation}
    \begin{gathered}
        \eta_{i,1}|\lambda_i=r\sim \text{Multinomial}(\pi_1^{(r)},\ldots,\pi_L^{(r)}), r=1,\ldots,R, i=1,\ldots,N; \\
        \boldsymbol{\pi}^{(1)}=(0.142, 0.137, 0.299, 0.422), \boldsymbol{\pi}^{(2)}=(0.326, 0.148, 0.127, 0.399), \\
        \boldsymbol{\pi}^{(3)}=(0.131, 0.282, 0.449, 0.138).
    \end{gathered}\label{eq:4.3.1}
\end{equation}
and the transition of latent classes at $t=2,\ldots,T$ follows:
\begin{equation}
    \begin{gathered}
        \eta_{i,t+1}|\eta_{i,t}=l,\lambda_i=r\sim \text{Multinomial}(\tau_{l,1}^{(r)},\ldots,\tau_{l,L}^{(r)}), r = 1,\ldots,R, l=1,\ldots,L, t=1,\ldots,T-1; \\
        \boldsymbol{\tau}^{(1)}=\begin{pmatrix} 0.965 & 0.010 & 0.014 & 0.011 \\ 0.007 & 0.226 & 0.166 & 0.601 \\ 0.028 & 0.130 & 0.318 & 0.524 \\ 0.002 & 0.126 & 0.182 & 0.691 \end{pmatrix}, \boldsymbol{\tau}^{(2)}=\begin{pmatrix} 0.577 & 0.190 & 0.227 & 0.005 \\ 0.315 & 0.458 & 0.215 & 0.012 \\ 0.341 & 0.256 & 0.394 & 0.009 \\ 0.005 & 0.016 & 0.009 & 0.970 \end{pmatrix}, \\
        \boldsymbol{\tau}^{(3)}=\begin{pmatrix} 0.109 & 0.257 & 0.440 & 0.195 \\ 0.063 & 0.784 & 0.032 & 0.121 \\ 0.095 & 0.075 & 0.745 & 0.085 \\ 0.088 & 0.356 & 0.233 & 0.323 \end{pmatrix}.
    \end{gathered}\label{eq:4.3.2}
\end{equation}

After generating the latent class sequences, the longitudinal item-responses are generated accordingly through $x_{itk}|\eta_{i,t}=l\sim \text{Multinomial}(p_{l1}^{(k)}, \ldots, p_{ld_k}^{(k)})$ as the second step, where the number of categories $d_k$ takes values in $\{2,5,6\}$, $k=1,\ldots,K$. The illustration of the latent class-specific item-response probabilities can be found in Figure S3. The simulated individual latent class sequences and their transitions can be found in Figure S4 and S5. Finally, to better reflect the real world settings, we introduce missingness at both the visit and item levels. For each individual, we randomly pick 0-4 visits to be missing, and for each non-missing individual visit, 0-10 items are further set to be missing.

We consider two types of outcomes in this simulation experiment: a binary outcome and a nominal outcome. The binary outcome is generated as 
\begin{equation}
    \begin{gathered}
        Y_i\sim \text{Bernoulli}(\theta_i), \\
        \text{logit}(\theta_i)=\beta_0+\beta_1\mathbbm{1}(\eta_{i,T}=2)+\beta_2\mathbbm{1}(\eta_{i,T}=3)+\beta_3\mathbbm{1}(\eta_{i,T}=4)+\beta_4\mathbbm{1}(\lambda_i=2)+\beta_5\mathbbm{1}(\lambda_i=3), \\
        \boldsymbol{\beta}=(-1.15, 0.95, 0.35, -0.70, 0.60, -0.55).
    \end{gathered}
\end{equation}
and is subsequently fitted through a logistic regression model. For the nominal outcome, we generate a three-category outcome that takes values in 0, 1 and 2 with category 0 as the reference category:
\begin{equation}
    \begin{gathered}
        Y_i\sim \text{Multinomial}(1;q_{i0},q_{i1},q_{i2}), \\
        \log(q_{ij}/q_{i0})=\gamma_{j0}+\gamma_{j1}\mathbbm{1}(\eta_{i,T}=2)+\gamma_{j2}\mathbbm{1}(\eta_{i,T}=3)+\gamma_{j3}\mathbbm{1}(\eta_{i,T}=4)+\gamma_{j4}\mathbbm{1}(\lambda_i=2)+\gamma_{j5}\mathbbm{1}(\lambda_i=3), \\
        \boldsymbol{\gamma}_1=(-2.10, 0.65, 0.95, 1.20, 0.50, 0.80), \boldsymbol{\gamma}_2=(-2.45, 0.35, 0.60, 0.90, 0.25, 0.50).
    \end{gathered}
\end{equation}
The nominal outcome is then modeled through a multinomial logistic regression model. The simulated binary outcome has an average event rate of 29.2\% across $S=200$ simulations, while for the nominal outcome, the average proportions of each category are 61.5\%, 27.2\% and 11.3\%, respectively.

\subsection{Method Evaluation}\label{sim:evaluation}

To evaluate the performance of the proposed method and demonstrate its superiority over two-stage alternatives, the simulated data are analyzed using both the joint model and a two-stage approach. The joint model follows the framework described in Section \ref{sec:method}. For the two-stage approach, we first model the longitudinal item-responses as described in Section \ref{joint:longitudinal}, and obtain posterior probabilities for each individual's sequence cluster membership and latent class memberships. The sequence cluster membership for individual $i$ is then assigned to the cluster with the highest posterior probability: $\hat{\lambda}_i=\max_r \hat{\text{Pr}}(\lambda_i=r\mid \boldsymbol{x}_i)$, and the latent class memberships over time are assigned analogously. In the second stage, the estimated $\hat{\lambda}_i$ and $\hat{\eta}_{i,T}$ are used as covariates in the outcome regression models.

We evaluate the performance of the methods based on coverage rate (the 95\% credible interval is formed via empirical percentiles from the draws), bias, average interval length, and root mean square error (RMSE).
In addition, we assess how well our method recovers the true class assignments by measuring the agreement between the true and estimated class memberships using the adjusted Rand index (ARI). We conduct 200 simulations.

\begin{figure}
\centerline{\includegraphics[scale=0.5]{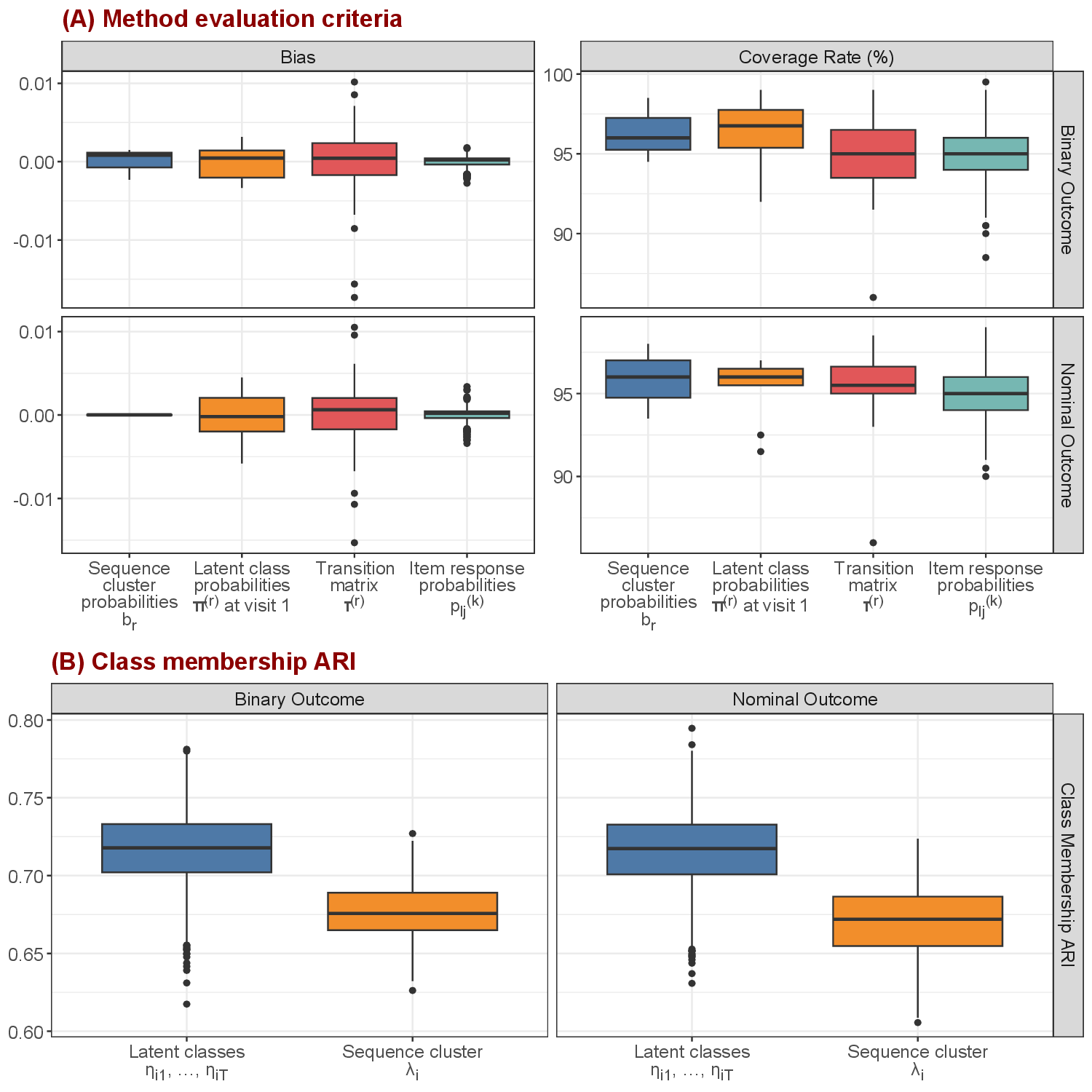}}
\caption{Simulation results for parameters associated with the longitudinal item-responses estimated from the joint model. Panel (A) summarizes the model performance regarding bias and coverage rate (\%) for parameters $b_r$, $\boldsymbol{\pi}^{(r)}$, $\boldsymbol{\tau}^{(r)}, r=1,\ldots,R$, and $p_{lj}^{(k)},l=1,\ldots,L,j=1,\ldots,d_k,k=1,\ldots,K$. Panel (B) presents the the adjusted Rand index (ARI) for evaluating the agreement between the true and estimated individual latent class memberships over time and sequence cluster assignments.} \label{fig:sim_result}
\end{figure}

Table \ref{tab:sim_result_table} summarizes the simulation results for the outcome parameters, comparing the proposed joint model with the two-stage alternative. Figure \ref{fig:sim_result} presents the results for the parameters associated with the longitudinal item-responses estimated from the joint model. From both the table and the figure, the joint model produces consistently unbiased estimates with desirable coverage rates across different outcome types. In addition, it is able to recover the true latent class and sequence cluster memberships as reflected by relatively high ARI values. In contrast, while the two-stage approach yields reasonably acceptable coverage for the binary outcome coefficients, it performs substantially worse for the nominal outcomes. Moreover, estimates from the two-stage approach exhibit non-negligible bias and are consistently attenuated towards the null. Overall, these simulation studies demonstrate the effectiveness of the proposed joint model.

\begin{table}
\caption{Simulation results including coverage rate, bias, average interval length and RMSE of outcome coefficients comparing joint model with two-stage approach. The first panel on the left presents the results of the outcome coefficients corresponding to the simulated binary outcomes. The two panels on the right present the results corresponding to the simulated nominal results. }
\label{tab:sim_result_table}
\begin{center}
\renewcommand{\arraystretch}{1.2}
\begin{threeparttable}
\begin{adjustbox}{width=\textwidth}

\begin{tabular}{l
                lcccc
                lcccc
                lcccc}
\toprule

& \multicolumn{5}{c}{Binary Outcomes}
& \multicolumn{5}{c}{Nominal Outcomes: 1 vs 0}
& \multicolumn{5}{c}{Nominal Outcomes: 2 vs 0} \\
\cmidrule(lr){2-6}\cmidrule(lr){7-11}\cmidrule(lr){12-16}
Method & Parameter & Coverage (\%) & Bias & \shortstack{Average\\Interval Length} & RMSE
& Parameter & Coverage (\%) & Bias & \shortstack{Average\\Interval Length} & RMSE
& Parameter & Coverage (\%) & Bias & \shortstack{Average\\Interval Length} & RMSE \\
\midrule

 & $\beta_0=-1.15$ & 95.5 & -0.00 & 0.96 & 0.24 & $\gamma_{10}=-2.10$ & 95.5 & -0.01 & 1.04 & 0.26 & $\gamma_{20}=-2.45$ & 91.0 & -0.03 & 1.35 & 0.37 \\
& $\beta_1=0.95$ & 96.0 & -0.01 & 0.88 & 0.23 & $\gamma_{11}=0.65$ & 96.0 & -0.02 & 0.99 & 0.26 & $\gamma_{21}=0.35$ & 95.0 & -0.02 & 1.34 & 0.35 \\
Joint Model & $\beta_2=0.35$ & 94.5 & -0.00 & 0.90 & 0.25 & $\gamma_{12}=0.95$ & 94.5 & -0.02 & 0.90 & 0.26 & $\gamma_{22}=0.60$ & 94.5 & -0.02 & 1.22 & 0.30 \\
& $\beta_3=-0.70$ & 95.0 & 0.00 & 0.88 & 0.23 & $\gamma_{13}=1.20$ & 95.0 & -0.01 & 0.86 & 0.22 & $\gamma_{23}=0.90$ & 94.5 & -0.00 & 1.13 & 0.29 \\
& $\beta_4=0.60$ & 94.5 & -0.00 & 0.83 & 0.20 & $\gamma_{14}=0.50$ & 95.5 & -0.01 & 0.89 & 0.22 & $\gamma_{24}=0.25$ & 93.5 & 0.02 & 1.17 & 0.31 \\
& $\beta_5=-0.55$ & 96.5 & -0.01 & 0.93 & 0.22 & $\gamma_{15}=0.80$ & 96.5 & 0.01 & 0.93 & 0.23 & $\gamma_{25}=0.50$ & 92.0 & 0.01 & 1.22 & 0.34 \\
\addlinespace[10pt]
& $\beta_0=-1.15$ & 93.5 & 0.08 & 0.78 & 0.23 & $\gamma_{10}=-2.10$ & 64.0 & 0.37 & 0.82 & 0.50 & $\gamma_{20}=-2.45$ & 85.5 & 0.23 & 1.10 & 0.48 \\
& $\beta_1=0.95$ & 90.0 & -0.08 & 0.74 & 0.22 & $\gamma_{11}=0.65$ & 80.0 & -0.23 & 0.83 & 0.33 & $\gamma_{21}=0.35$ & 91.5 & -0.17 & 1.17 & 0.35 \\
Two-stage approach & $\beta_2=0.35$ & 85.0 & -0.14 & 0.66 & 0.23 & $\gamma_{12}=0.95$ & 56.5 & -0.30 & 0.67 & 0.37 & $\gamma_{22}=-0.70$ & 91.0 & 0.04 & 0.74 & 0.26 \\
& $\beta_3=-0.70$ & 91.0 & 0.04 & 0.74 & 0.22 & $\gamma_{13}=1.20$ & 63.5 & -0.26 & 0.71 & 0.36 & $\gamma_{23}=0.90$ & 83.5 & -0.15 & 0.97 & 0.39 \\
& $\beta_4=0.60$ & 88.0 & -0.16 & 0.69 & 0.24 & $\gamma_{14}=0.50$ & 90.0 & -0.14 & 0.71 & 0.28 & $\gamma_{24}=0.25$ & 91.5 & -0.15 & 0.96 & 0.65 \\
& $\beta_5=-0.55$ & 89.5 & 0.11 & 0.75 & 0.23 & $\gamma_{15}=0.80$ & 80.5 & -0.30 & 0.75 & 0.76 & $\gamma_{25}=0.50$ & 89.0 & -0.24 & 1.01 & 0.85 \\
\bottomrule
\end{tabular}

\end{adjustbox}
\end{threeparttable}
\end{center}
\end{table}

\section{Application to the Longitudinal Symptomatology Scale Responses from SWAN}\label{sec:application}

\subsection{Study Description}\label{app:intro}

The Study of Women's Health Across the Nation (SWAN) is a multi-site, multi-racial, and multi-ethnic longitudinal cohort study that follows midlife women through the menopausal transition \citep{swan}. Eligible participants were women aged 42-52 years who had a uterus and at least one ovary, were not on exogenous hormones, had at least one menstrual period in the past three months, and self-identified with one of the predesignated race or ethnic groups. 3302 females were enrolled in 1996 and have been followed annually for approximately 16 visits. The annual visits include questionnaires, blood and urine specimen collection and physical measures. To investigate factors impacting the menopausal transition, the data collection instruments and approaches involve biological, psychological/psychosocial, sociocultural/environmental, and feminist approaches. Detailed descriptions of the study design and data collection can be found in \citet{swan}. In this application study, we focus on participants' responses to the longitudinal symptomatology scales and examine how individual symptom profiles influence subsequent health outcomes. 

The symptomatology scales capture a broad range of physical and psychological symptoms and include the 36-item Short-Form Health Survey (SF-36, \citet{ware1993sf}), the Pittsburgh Sleep Quality Index (PSQI, \citet{buysse1989pittsburgh}), the Center for Epidemiologic Studies Depression Scale (CES-D, \citet{radloff1977ces}), the 4-item Cohen's Perceived Stress Scale \citep{cohen1983global}, and 19 miscellaneous items around sleep, menopause-related symptoms and mood. Together, these scales contain 98 items, though some items are administered only at certain visits. To maximize the number of items while maintaining measurement consistency across visits, we focus on 60 items collected at every other visit from baseline to visit 12 (7 visits in total), including SF-36 domains of role-physical, bodily pain, vitality, social functioning and role-emotional, one item from SF-36 domain of general health, an overall sleep quality item from PSQI, the full CES-D and Cohen's Perceived Stress Scale, and the 19 miscellaneous items. 
Based on individual's longitudinal response patterns, we aim to classify their responses at each visit into latent classes, group the resulting latent classes sequences into sequence clusters, and investigate how these latent profiles are associated with subsequent health outcomes. 

The health outcome of interest is the fall patterns collected at visit 15. Participants were asked whether they had fallen and landed on the floor or ground (or fallen and hit an object like a table or stair) in the past year and, if so, how many times. We aim to examine whether individuals' latent class membership at visit 12 and their sequence cluster membership are associated with (a) the occurrence of any fall and (b) fall frequency, categorized as no falls, a single fall or recurrent falls. Baseline BMI was included as a covariate. After excluding females (a) without responses to the fall pattern questions at visit 15; (b) with more than four missing visits; (c) with baseline BMI missing, and removing individual visits with more than 20 items without responses, our final analytical sample contains 1934 females with an average of 6.2 visits and 61.2 non-missing item responses per visit. Among these females, 1360 reported no falls in the past year at visit 15, 328 reported a single fall, and 246 reported recurrent falls. The average baseline BMI was 27.61 (SD = 7.07).

\subsection{Method Implementation}\label{app:method}

To accomplish our study objective, we fit two separate joint models. In the first model, we investigate the association between longitudinal patterns of item-responses and the occurrence of any fall in the past year, and the outcome is connected with the longitudinal component through a logistic regression model. In the second model, the fall frequency categories are associated with the longitudinal patterns through a multinomial logistic regression model. For both models, the individual sequence clusters $\lambda_i$, their latent class memberships at the 12th annual visit $\eta_{i,T}$, and their baseline BMI measurement are used as predictors for the outcome submodel.

\subsection{Results}

Guided by a previous study \citep{harlow2017not}, which analyzed a similar set of longitudinal symptomatology data from SWAN, and using AIC and BIC as model selection criteria (the detailed model selection criteria can be found in Supplementary Materials Section 2.2), our method identifies 6 item-response latent classes for both the fall occurrence model and the fall frequency model consistent with the number of classes in \cite{harlow2017not}.  We also identify 4 distinct sequence clusters corresponding to different baseline distributions and transition probabilities for the latent classes.

\subsubsection{Outcome I: Fall Occurrence}\label{app:outcomeI}

\paragraph{Item-Response Latent Classes}\label{outcomeI:LC}

\begin{figure}
\centerline{\includegraphics[scale=0.5]{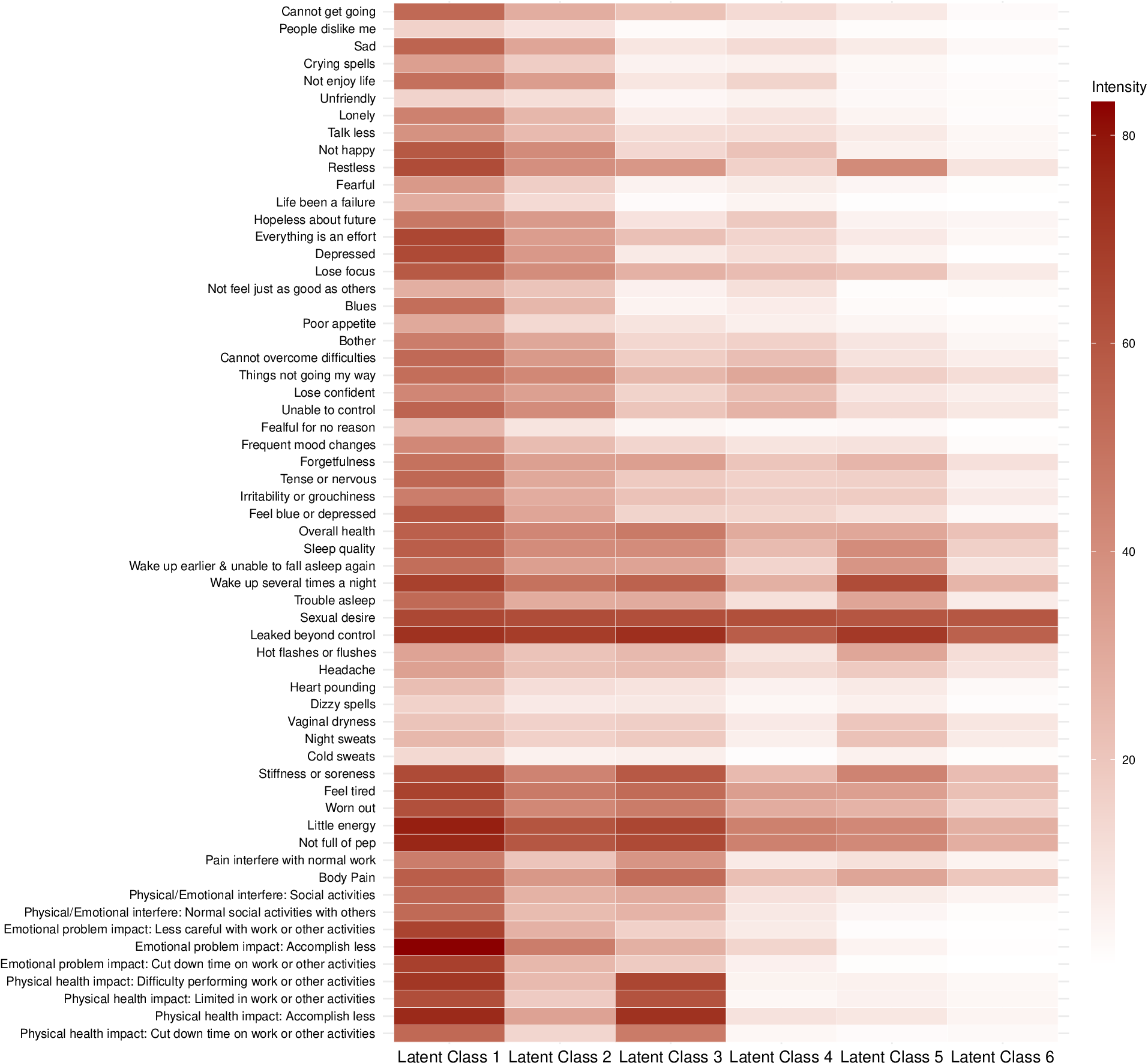}}
\caption{Heatmap of symptom intensity by the six identified item-response latent classes from the joint model with outcome of fall occurrence, with darker red representing higher intensity. From left to right, the latent classes are ordered from class with most symptoms to class with least symptoms. The names on the left are shorter labels for the originally administered items.} \label{fig:swan_intensity_fall}
\end{figure}

Figure \ref{fig:swan_intensity_fall} presents the symptom intensity of the six identified latent classes using a heatmap representation. The average symptom intensity for the $k$-th item in the $l$-th latent class is calculated by $\sum_{j=1}^{d_k}p_{lj}^{(k)}(j-1)/(d_k-1)$ on a scale of [0,1], with values closer to 1 indicating greater symptom severity. The latent classes are ordered by decreasing overall symptom intensity. Latent class 1 exhibits high symptom intensity across most items and is labeled as the ``high symptom class", whereas latent class 2 displays a similar but less severe profile and is labeled the ``moderate symptom class". Unlike latent class 1 and 2, latent class 3 is characterized by moderate to high intensity in physical symptoms, including impact of physical health, bodily pain, loss of vitality and sleep problems, and is therefore referred to as the ``high physical symptoms class". The remaining classes represent relatively healthier profiles. We label latent class 4 as the ``fatigue \& moderate psychological symptoms class" based on the moderate symptom intensity with respect to loss of vitality and depressive symptomatology items. Latent class 5 exhibits fewer psychological issues but more pronounced fatigue and sleep problems, and we name it as the ``fatigue class". Latent class 6 has the lowest overall symptom intensity, with only mild fatigue-related symptoms, and is labeled as the ``low symptoms \& fatigue class". Notably, high symptom intensity with respect to leaked beyond control and lose of sexual desire is observed across all latent classes. (Note the similarity between these LCs and the LCs defined in \cite{harlow2017not}, which used a similar, though not identical, set of symptoms in their latent transition analysis.) In the following text, we refer to the latent classes as LC 1-6 and their assigned names interchangeably.

\paragraph{Sequence Cluster Profile}\label{outcomeI:SC}

\begin{figure}
\centerline{\includegraphics[scale=0.33]{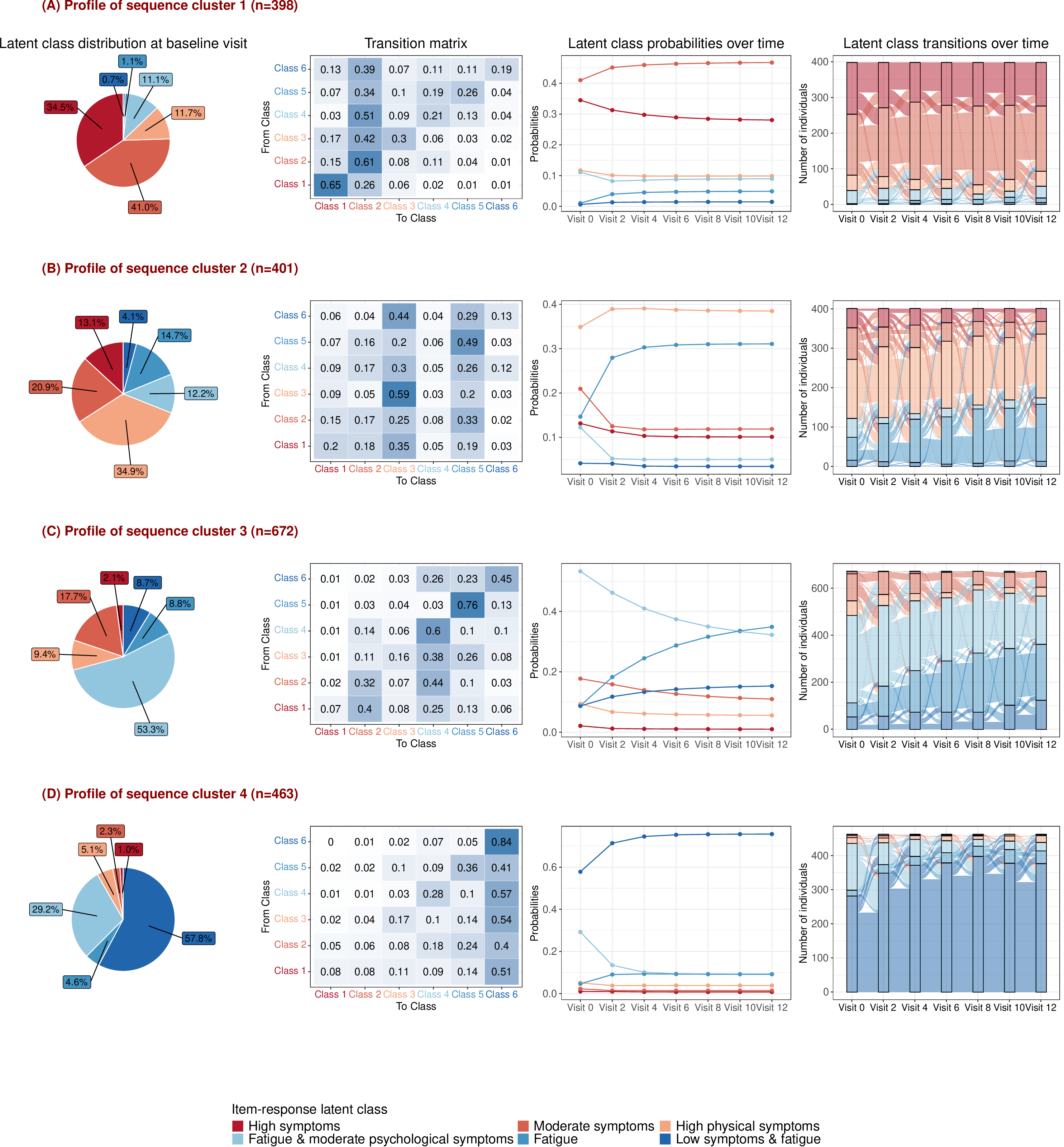}}
\captionsetup{font=footnotesize}
\caption{Profile of the four identified sequence clusters from the joint model with outcome of fall occurrence. Each row presents the profile of one sequence cluster. The panels from left to right on the $r$-th row are: a pie chart representation of initial probabilities $\boldsymbol{\pi}^{(r)}$; the transition matrix $\boldsymbol{\tau}^{(r)}$, with darker blue indicating higher transition probabilities; a trajectory plot illustrating the development of latent class probabilities across visits, i.e., a representation of $\hat{\text{Pr}}(\eta_{it}=l|\lambda_i=r), l=1,\ldots,R, t = 1,\ldots,T$; and an alluvial diagram visualizing individual-level latent class transitions over time, with the individual latent class membership assigned based on modal class assignment.} \label{fig:swan_profile_fall}
\end{figure}

Figure \ref{fig:swan_profile_fall} presents the profiles of the four identified sequence clusters, summarized by $\boldsymbol{\pi}^{(r)}$, $\boldsymbol{\tau}^{(r)}$ and $\hat{\text{Pr}}(\eta_{i,t}=l|\lambda_i), l = 1,\ldots,L, t = 1,\ldots,T$, with the clusters ordered by increasing overall health status. Sequence cluster 1 contains 20\% of the sample and is characterized by persistently poor health. At the baseline visit, most females belong to either the high symptoms class (34.5\%) or the moderate symptoms class (41.0\%). Over time, females in the high and moderate symptoms classes tend to remain in the same class (61.2\% and 65.1\%) or transition between each other (15.4\% and 25.6\%). In contrast, females in the less symptomatic LC3-LC6 exhibit only moderate probabilities of staying in their own classes (18.7\%-29.9\%) and are more likely to transition to the moderate symptoms class (34.2\%-51.0\%) than to healthier classes. The trajectory and alluvial plots further show that the moderate and high symptoms classes consistently dominate this cluster throughout follow-up, indicating persistent or worsening symptoms over time and the poorest overall health status among the four sequence clusters. 

Sequence cluster 2 consists of 21\% of the sample. At baseline visit, more than half of the females fall into the moderate symptoms class (20.9\%) and the high physical symptoms class (34.9\%), with smaller but comparable proportions in the high symptoms class (13.1\%), the fatigue \& moderate psychological symptoms class (12.2\%) and the fatigue class (14.7\%). The transition matrix indicates a strong tendency to remain in or transition to the high physical symptoms class and the fatigue class. Females in these two classes are relatively stable, with a high probability of remaining in the same class (59.3\% and 49.2\%), and there are also noticeable transitions between them (19.8\% and 19.6\%). For females in the remaining classes, transitions to the high physical symptoms class (25.1\%-44.4\%) and the fatigue class (19.2\%-33.0\%) are common, while transitions to other classes, particularly LC4 and LC6, are relatively uncommon. The trajectory and alluvial plots further demonstrate a consistently high proportion of females in the high physical symptoms class and an increasing proportion in the fatigue class over time, suggesting a general shift toward classes dominated by physical burden and fatigue symptoms. 

Sequence cluster 3 includes 35\% of the sample and is characterized by a predominance of the fatigue \& moderate psychological symptoms class (53.3\%) at the baseline visit. The remaining females are distributed relatively evenly across the other classes, except for a very small proportion in the high symptoms class. Females in LC1-LC3 tend to transition toward less symptomatic classes or remain in their current classes, showing little evidence of progression to more severe symptom classes. Females in LC4 and LC5 exhibit relatively high stability, with probabilities of staying in the same class of 59.5\% and 76.1\%. Females in LC6 also exhibit moderate stability (45.2\%), with some developing more pronounced fatigue symptoms and moderate psychological symptoms. This cluster is overall characterized by a gradual decline in the proportion of females in the fatigue \& moderate psychological symptoms class, accomplished by a steady increase in the fatigue class. The overall health status of females in sequence cluster 3 is much better compared to the first two clusters.

The remaining 24\% of the sample belong to sequence cluster 4, the healthiest group among the four sequence clusters. At the baseline visit, this cluster is primarily composed of females in the low symptoms \& fatigue class (57.8\%) and the fatigue \& moderate psychological symptoms class (29.2\%). Over time, females tend to transition to or remain in classes with less symptoms regardless of their initial class membership. In particular, the low symptoms \& fatigue class acts as an absorbing state and becomes increasingly prevalent over time. The trajectory and alluvial plots further suggest a consistent improvement or maintenance of better health status throughout follow-up.

\paragraph{Association with Fall Occurrence}\label{outcomeI:results}

Panel (A) in Table \ref{tab:odds_ratio} summarizes the outcome model results in terms of posterior mean odds ratio. Compared with individuals in sequence cluster 1, those in other sequence clusters do not exhibit significantly different odds of falling. However, further pairwise comparisons reveal that individuals in sequence cluster 2 have significantly higher odds of falling than those in sequence cluster 3 (OR = 1.82, 95\% CrI: 1.15, 2.76) and sequence cluster 4 (OR = 1.92, 95\% CrI: 1.09, 3.16). These findings indicate that individuals with a strong tendency to transition toward and remain in latent classes with high physical burden (LC 3) and fatigue symptoms (LC 5) have nearly twice the risk of falling compared with individuals who follow healthier symptom trajectories. 

\begin{table}
\caption{Association between individual sequence cluster memberships and latent class memberships at the 12th visit and fall outcomes at Visit 15. Panel (A) presents odds ratio and 95\% credible intervals (CrI) for the fall occurrence outcome and Panel (B) presents the odds ratio and 95\% credible intervals (CrI) for the fall frequency outcome.}
\label{tab:odds_ratio}
\renewcommand{\arraystretch}{1.2}
\begin{center}
\begin{threeparttable}
\begin{adjustbox}{width=\textwidth}

\begin{tabular}{lccc}
\toprule

& \multicolumn{1}{c}{(A) Outcome I: fall occurrence}
& \multicolumn{2}{c}{(B): Outcome II: fall frequency}\\
& & \multicolumn{1}{c}{A single fall vs never fall} & \multicolumn{1}{c}{Recurrent falls vs never fall}\\
\cmidrule(lr){3-4}
Covariate & Odds Ratio (95\% CrI) & Odds Ratio (95\% CrI) & Odds Ratio (95\% CrI)\\
\midrule

Sequence cluster (ref = sequence cluster 1) & & & \\
Sequence cluster 2 & 1.59 (0.92, 2.61) & 1.32 (0.68, 2.31) & 1.98 (0.99, 3.57) \\
Sequence cluster 3 & 0.89 (0.53, 1.43) & 1.00 (0.51, 1.77) & 0.77 (0.37, 1.42) \\
Sequence cluster 4 & 0.87 (0.47, 1.54) & 0.73 (0.34, 1.40) & 1.18 (0.47, 2.45) \\
\addlinespace[7pt]
Latent class at visit 12 (ref = High symptoms class) & & & \\
Moderate symptoms class & \textbf{0.64 (0.41, 0.96)} & 0.85 (0.47, 1.40) & \textbf{0.57 (0.31, 0.95)} \\
High physical symptoms class & \textbf{0.41 (0.23, 0.67)} & 0.67 (0.34, 1.19) & \textbf{0.29 (0.14, 0.51)} \\
Fatigue \& moderate psychological symptoms class & \textbf{0.45 (0.24, 0.74)} & 0.66 (0.32, 1.22) & \textbf{0.34 (0.14, 0.66)} \\
Fatigue class & \textbf{0.57 (0.33, 0.91)} & 0.77 (0.39, 1.38) & \textbf{0.52 (0.27, 0.91)} \\
Low symptoms class & \textbf{0.38 (0.20, 0.66)} & 0.77 (0.35, 1.43) & \textbf{0.17 (0.07, 0.35)} \\
\addlinespace[7pt]
Baseline BMI (by 5 units) & 1.02 (0.95, 1.10) & 0.96 (0.88, 1.06) & 1.09 (0.99, 1.19)\\
\bottomrule
\end{tabular}

\end{adjustbox}
\end{threeparttable}
\end{center}
\end{table}

Latent class membership at visit 12, the closet visit included in the analysis to visit 15, also shows a strong association with fall occurrence. Compared with individuals in the high symptoms class (LC1), those in LC2-LC6 have significantly lower odds of falling. For example, individuals in LC2 have 64.1\% (95\% CrI: 40.8\%, 95.7\%) of the odds of falling relative to those in LC1, while individuals in LC6 have the lowest risk, with only 37.9\% (95\% CrI: 19.5\%, 65.8\%) of the odds of falling. These findings suggest that fall occurrence is also strongly driven by an individual's most recent symptom profile. In addition, individuals in the high and moderate symptoms classes have higher odds of falling than those in the high physical symptoms class, indicating that beyond physical symptoms, psychological symptom burden may further increase fall risk. Baseline BMI is not significantly associated with fall occurrence, with a five-unit increase in baseline BMI associated with a non-significant 2.03\% (95\% CrI: -5.25\%, 9.19\%) increase in the odds of falling.

\subsubsection{Outcome II: Fall Frequency}\label{app:outcomeII}

\paragraph{Item-Response Latent Classes and Sequence Cluster Profile}\label{outcomeII:profile}

The profiles of the six identified item-response latent classes and the four identified sequence clusters can be found in Figure S6 and Figure S7 in Supplementary Materials, respectively. These profiles closely resemble those obtained from the fall occurrence outcome model, which is expected since the two outcomes are both defined base on the same underlying measure of falls. As a result, we keep the same labels and interpretations for the latent classes and sequence clusters as discussed in \ref{outcomeI:LC} and \ref{outcomeI:SC} in the following text.

\paragraph{Association with Fall Frequency}

Panel (B) of Table \ref{tab:odds_ratio} presents the associations between individual sequence cluster and latent class membership at visit 12 and fall frequency, reported as posterior mean odds ratio. Studying the fall frequency can help us further understand how the longitudinal item-response patterns affect the occurrence of fall. From the first column of Panel (B), sequence cluster membership is not significantly associated with the odds of experiencing a single fall relative to no falls. Similarly, latent class membership at visit 12 does not appear to significantly change individuals' odds of having a single fall relative to no falls. Baseline BMI also does not exhibit a significant association with individuals' odds of having a single fall over no falls. These findings suggest that the occurrence of a single fall is more of a random event and is likely not driven by the long-term symptom trajectories, the most recent symptom profiles or the baseline characteristics.

The second column of Panel (B) in Table \ref{tab:odds_ratio} gives us more insights into the risk factors of the falls. First, latent class membership at visit 12 shows a strong association with the odds of recurrent falls. Individuals in the high symptoms class have the highest risk, while those in the remaining classes have significantly lower odds of recurrent falls. For example, compared to individuals in the high symptoms class, those in the moderate symptoms class have 43.2\% (95\% CrI: 5.3\%, 68.7\%) lower odds of experiencing recurrent falls relative to no falls, and those in the low symptoms \& fatigue class have the lowest risk to experience recurrent falls (OR = 0.17, 95\% CrI: 0.07, 0.35). Comparisons between the high and moderate symptoms classes and the high physical symptoms class further emphasize that recurrent falls may be a joint consequence of physical and psychological symptoms.

The symptom trajectories, quantified by the sequence cluster, also explains the fall occurrence. Individuals in sequence cluster 2 have nearly twice the odds of experiencing recurrent falls relative to no falls compared with those in sequence cluster 1 (OR = 1.98, 95\% CrI: 0.99, 3.57) and sequence cluster 3 (OR = 2.71, 95\% CrI: 1.42, 4.77). These findings suggest that individuals who tend to develop and retain high physical symptoms (LC3) and fatigue symptoms (LC5) are at particularly high risk of recurrent falls, even compared with individuals whose trajectories are consistently dominated by high and moderate symptoms. This pattern further suggests that, among individuals in sequence cluster 1, recurrent fall risk may be largely driven by the recent symptom profiles rather than the historical trajectories. The latter finding explains how individuals in sequence cluster 2 have higher odds of fall occurrence than individuals in sequence cluster 3 as described in \ref{outcomeI:results} through having higher odds of recurrent falls. Baseline BMI also has a meaningful, though non-significant, association with recurrent falls. With a five-unit increase in the baseline BMI, the odds of recurrent falls relative to no falls are increased by 9.1\% (95\% CrI: -0.7\%, 19.3\%).

\section{Discussion}
\label{sec:discussion}
The application of our method to the longitudinal symptomatology scale responses and self-reported fall outcomes from the SWAN study both confirms 
prior findings and provides additional insights into fall risk among midlife females:

\begin{itemize}
    \item We identify six latent symptom classes, including two classes with high or moderate intensity across all measured symptoms, one dominated by high physical symptoms, two characterized by a predominant fatigue burden with or without additional psychological symptoms, and one with generally low symptom intensity. These findings are consistent with a previous study by \citet{harlow2017not}, who examined a similar set of symptoms over menopausal transition (pre, early peri-, late peri- and post-menopause) and reported similar symptom subgroups.
    \item At the longitudinal level, we identified four distinct sequence clusters. The first cluster is characterized by persistent high and moderate symptom burden. The second begins with a relatively even distribution across symptom classes but exhibits a strong tendency to transition toward and remain in classes dominated by physical symptoms and fatigue. The third is primarily composed of females in less symptomatic classes at baseline, with substantial stability in less symptomatic classes and transitions from more symptomatic classes to less symptomatic states. The last one is dominated by low or fatigue symptoms at baseline and demonstrates consistent transitions toward healthier profiles over time. These patterns highlight substantial heterogeneity in symptom evolution during midlife, suggesting that jointly monitoring symptom profiles and transition dynamics may help identify females at elevated risk of falls and facilitate targeted interventions.
    \item Connecting the symptom classes and sequence clusters with fall outcomes confirms findings from previous literature and provides some additional insights. Females in the high and moderate symptom classes exhibit significantly higher odds of recurrent falls, supporting prior evidence that falls arise from a complex combination of symptoms from multiple domains \citep{white2018fall, peeters2019comprehensive}. In addition, individuals in the sequence cluster with a strong tendency to develop and retain high physical and fatigue symptoms over time have the highest risk of experiencing recurrent falls, indicating the importance of monitoring symptom trajectories rather than relying solely on cross-sectional symptom profiles. In contrast, we do not observe meaningful differences in the odds of a single fall across symptom classes and sequence clusters. As pointed out by \citet{ karvonen2020midlife}, recurrent falls are more strongly driven by intrinsic factors of females, whereas a single fall may be more likely caused by environmental factors. That explains why different symptom classes and sequence clusters did not show differences in the risk of single falls but significant differences in risk of recurrent falls. Also, the fact that the most symptomatic of the HTLA clusters does not correspond to the highest risk for recurrent falls suggests that high symptomatology may lead to reduced mobility so that fall risk decreases, suggesting a ``U-shaped'' risk of falls relative to overall health status.
\end{itemize}

Several limitations of our study warrant discussion. First, we assume that the latent symptom classes are invariant across demographic and socioeconomic characteristics, which may be restrictive as previous work has shown that factors such as financial strain, obesity and smoking are associated with symptom class membership among midlife women \citet{harlow2017not}. Future work could incorporate baseline covariates when estimating symptom classes. Second, we assume time-invariant transition probabilities within sequence clusters, implying that symptom dynamics remain constant over time. This assumption may be problematic during menopausal transition that also occurs in midlife, which is accompanied by substantial physiological and hormonal changes. Extending the model to allow for time-varying or menopausal stage-specific transition probabilities would provide a more flexible representation of symptom evolution at the cost of reduced interpretability of the model. Finally, while our analysis focuses on fall outcomes at a single visit, SWAN has collected fall outcomes over multiple annual visits. Future work could extend the current framework to develop a dynamic prediction model that incorporates both longitudinal symptom trajectories and fall history. Such an extension could enable individualized risk prediction and support targeted interventions.

\section*{Acknowledgments}

The Study of Women's Health Across the Nation (SWAN) has grant support from the National Institutes of Health (NIH), DHHS, through the National Institute on Aging (NIA), the National Institute of Nursing Research (NINR) and the NIH Office of Research on Women’s Health (ORWH) (Grants U01NR004061; U01AG012505, U01AG012535, U01AG012531, U01AG012539, U01AG012546, U01AG012553, U01AG012554, U01AG012495, and U19AG063720). The content of this article is solely the responsibility of the authors and does not necessarily represent the official views of the NIA, NINR, ORWH or the NIH. This research was also partly supported through computational resources and services provided by Advanced Research Computing (ARC), a division of Information and Technology Services (ITS) at the University of Michigan, Ann Arbor.

Clinical Centers: \textit{University of Michigan, Ann Arbor---Carrie Karvonen-Gutierrez, PI 2021–present, Siobán Harlow, PI 2011–2021, MaryFran Sowers, PI 1994-2011; Massachusetts General Hospital, Boston, MA---Sherri‐Ann Burnett‐Bowie, PI 2020–Present; Joel Finkelstein, PI 1999–2020; Robert Neer, PI 1994–1999; Rush University, Rush University Medical Center, Chicago, IL---Imke Janssen, PI 2020–Present; Howard Kravitz, PI 2009–2020; Lynda Powell, PI 1994–2009; University of California, Davis/Kaiser---Elaine Waetjen and Monique Hedderson, PIs 2020–Present; Ellen Gold, PI 1994-2020; University of California, Los Angeles---Arun Karlamangla, PI 2020–Present; Gail Greendale, PI 1994-2020; Albert Einstein College of Medicine, Bronx, NY---Carol Derby, PI 2011–present, Rachel Wildman, PI 2010–2011; Nanette Santoro, PI 2004–2010; University of Medicine and Dentistry–New Jersey Medical School, Newark---Gerson Weiss, PI 1994–2004, and the University of Pittsburgh, Pittsburgh, PA---Rebecca Thurston, PI 2020–Present; Karen Matthews, PI 1994-2020}.

NIH Program Office: \textit{National Institute on Aging, Bethesda, MD---Rosaly Correa-de-Araujo 2020-present; Chhanda Dutta 2016-present; Winifred Rossi 2012–2016; Sherry Sherman 1994–2012; Marcia Ory 1994–2001; National Institute of Nursing Research, Bethesda, MD---Program Officers}.

Central Laboratory: \textit{University of Michigan, Ann Arbor---Daniel McConnell (Central Ligand Assay Satellite Services)}.

Coordinating Center: \textit{University of Pittsburgh, Pittsburgh, PA---Maria Mori Brooks, PI 2012-present; Kim Sutton-Tyrrell, PI 2001–2012; New England Research Institutes, Watertown, MA---Sonja McKinlay, PI 1995–2001}.

Steering Committee:	Susan Johnson, Current Chair, Chris Gallagher, Former Chair.

We thank the study staff at each site and all the women who participated in SWAN.

The authors would also like to thank the SWAN team at the University of Michigan for providing the datasets for analysis.

\section*{Disclosure Statement}

The authors report there are no competing interests to declare.

\section*{Data Availability Statement}\label{data-availability-statement}

The data that support findings in this paper will be shared on reasonable request.

\section*{Supplementary Materials}

The Supplementary Materials provide information on the relabeling algorithm, the model selection criteria and supplementary of the simulation study and application results.

\bibliographystyle{apalike}
\bibliography{references}  






\end{document}


\maketitle

\section{Relabel Algorithm}\label{method:relabel}

\begin{figure}
\begin{center}
\includegraphics[scale=0.75]{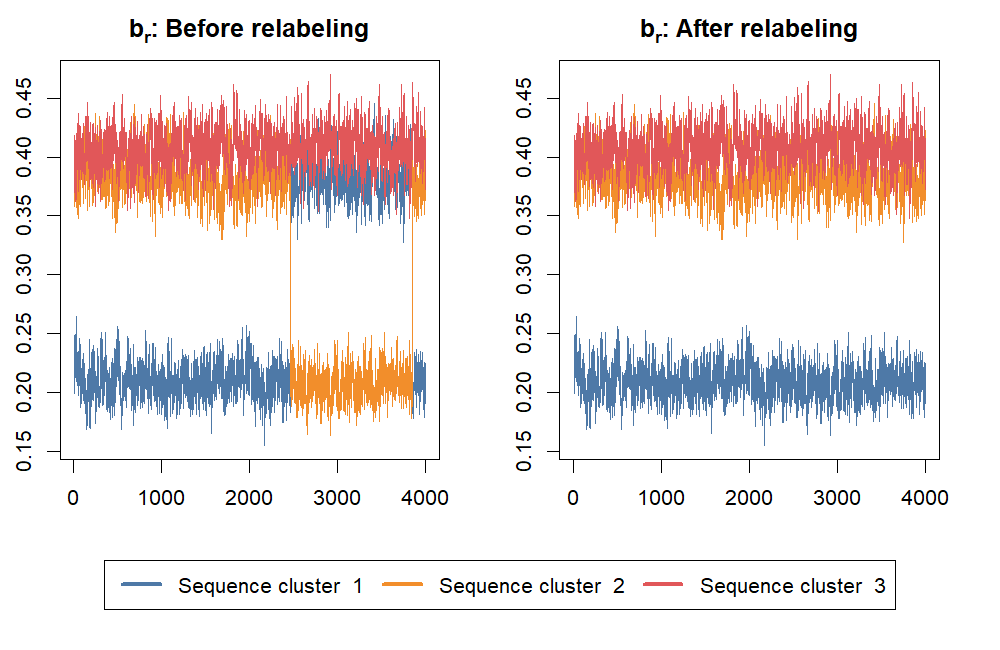}
\end{center}
\caption{The posterior draws of the sequence cluster probabilities $b_r, r=1,\ldots,3$, before and after applying the post-hoc relabeling algorithm from one simulation replication. \label{fig:relabel_br}}
\end{figure}

When Bayesian methods are used for parameter estimation and clustering in mixture models, label switching is a common issue. The root of label switching lies in the invariance of the data likelihood regardless of how we relabel the mixture components \citep{redner1984mixture}. Figure \ref{fig:relabel_br} shows an example of label switching when estimating $b_r=\text{Pr}(\lambda_i=r)$ in the simulation study in Section 3 of the main text, where the posterior draws jump between different regions. To address this issue, we adopt and extend Stephens' relabeling algorithm \citep{stephens2000dealing} to relabel the posterior samples. Suppose we are interested in classifying a study population of size $N$ into $S$ groups based on the observed data $\boldsymbol{x}$. Let $\boldsymbol{\theta}$ denote the model parameters, and let $Z_i$ denote the group membership of individual $i$, for $i=1,\ldots,N$. For the $m$-th posterior sample, $m=1,\ldots,M$, we can calculate an $N\times S$ matrix of classification probabilities $(u_{is}(\boldsymbol{\theta}^{(m)}))$, with $u_{is}(\boldsymbol{\theta}^{(m)})=\text{Pr}(Z_i=s\mid \boldsymbol{x}, \boldsymbol{\theta}^{(m)})$. In order to determine the group memberships, a natural way is to report an $N\times S$ posterior classification probability matrix $Q = (q_{is})$. In the absence of label switching, $q_{is}$ can be estimated by averaging $u_{is}(\boldsymbol{\theta}^{(m)})$ across $M$ posterior samples. However, due to the existence of label switching, the labels $\{1,\ldots,S\}$ may switch orders across iterations, making direct averaging invalid. The main idea of the Stephens' relabel algorithm is that, for each iteration, we apply a permutation $\nu_m$ to the labels and obtain an updated $\hat{q}_{is}$ estimate through averaging over the permuted quantities $u_{is}(\nu_1(\boldsymbol{\theta}^{(1)})), \ldots, u_{is}(\nu_M(\boldsymbol{\theta}^{(M)}))$. Given the current estimate $\hat{q}_{is}$, the algorithm try to update the permutations $\nu_1,\ldots,\nu_M$ of labels through minimizing a loss function. The above procedure is repeated until the permuted posterior samples agree on and yield a stable posterior classification probability matrix. Details of the Stephens' relabeling algorithm can be found in \citet{stephens2000dealing}. 

In our model setting, relabeling is required for both the labels of sequence clusters and the labels of the longitudinal item-response latent classes in the posterior samples. Let $\boldsymbol{\phi}$ denote the parameters in our model. To implement the relabeling algorithm, we need to compute the classification probabilities $\text{Pr}(\lambda_i=r\mid \boldsymbol{x}, \boldsymbol{y}, \boldsymbol{\phi})$ and $\text{Pr}(\eta_{i,t}=l\mid \boldsymbol{x}, \boldsymbol{y}, \boldsymbol{\phi})$ at each MCMC iteration. A straightforward way to do the computation, say for $\text{Pr}(\lambda_i=r\mid \boldsymbol{x}, \boldsymbol{y}, \boldsymbol{\phi})$, involves marginalizing over all possible latent class sequences $\{\eta_{i,1},\ldots,\eta_{i,T}\}$, i.e.,
$$\text{Pr}(\lambda_i=r\mid \boldsymbol{x}, \boldsymbol{y}, \boldsymbol{\phi}) = \sum_{l_1=1}^L\ldots \sum_{l_T=1}^L \text{Pr}(\lambda_i=r, \eta_{i,1}=l_1, \ldots, \eta_{i,T}=l_T\mid \boldsymbol{x}, \boldsymbol{y}, \boldsymbol{\phi}).$$ However, this direct computation becomes computationally prohibitive and memory demanding as $T$ increases. Therefore, we adopt a forward-backward algorithm to efficiently compute these marginal probabilities.   

\begin{figure}
\begin{center}
\includegraphics[scale=0.65]{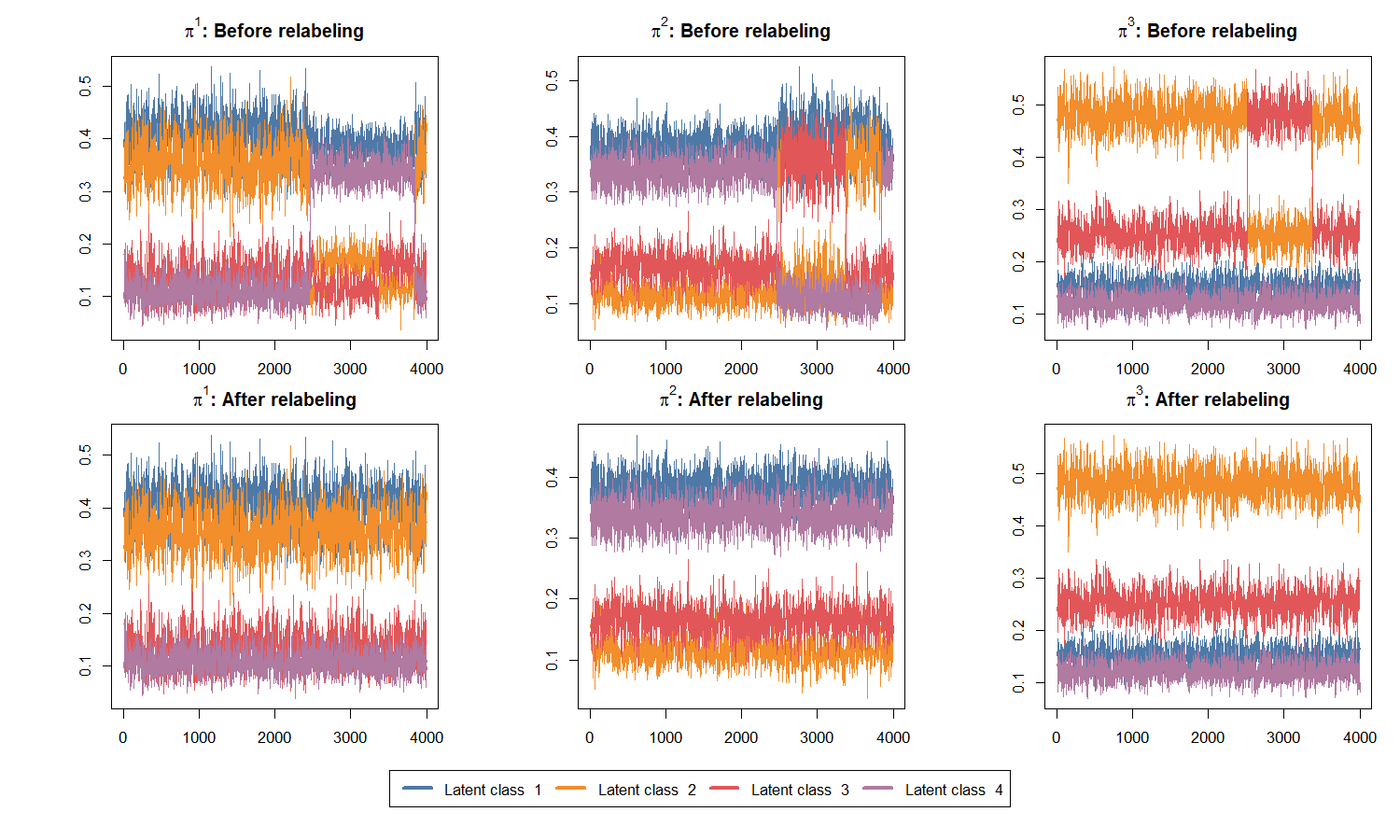}
\end{center}
\caption{The posterior draws of the latent class probabilities at the initial visit $\boldsymbol{\pi}^{(r)}, r=1,\ldots,3$, before and after applying the post-hoc relabeling algorithm from the same simulation replication as Figure \ref{fig:relabel_br}. \label{fig:relabel_pi}}
\end{figure}

The forward probabilities are defined as $\alpha_{i,t}^{(r)}(l)=\text{Pr}(\boldsymbol{x}_{i,1:t}, \eta_{i,t}=l\mid \lambda_i=r,\boldsymbol{\phi})$, with initialization at $t=1$ given by $\alpha_{i,1}^{(r)}(l)=\pi_l^{(r)}\text{Pr}(\boldsymbol{x}_{i,1}\mid \eta_{i,1}=l, \boldsymbol{\phi})$. For $t=2,\ldots,T$, the forward probabilities are computed recursively as $$\alpha_{i,t}^{(r)}(l)=\text{Pr}(\boldsymbol{x}_{i,t}\mid \eta_{i,t}=l, \boldsymbol{\phi})\sum_{l_\text{prev}=1}^L\tau_{l_{\text{prev}},l}^{(r)}\alpha_{i,t-1}^{(r)}(l_{\text{prev}}).$$ The backward probabilities are defined as $\beta_{i,t}^{(r)}(l)=\text{Pr}(y_i, \boldsymbol{x}_{i,t+1:T}\mid \eta_{i,t}=l, \lambda_i=r, \boldsymbol{\phi})$. They are initialized at $t=T$ as $\beta_{i,T}^{(r)}(l)=\text{Pr}(y_i\mid \eta_{i,T}=l,\lambda_i=r, \boldsymbol{\phi})$, which depends on the specified outcome model. For $t=T-1,\ldots,1$, $\beta_{i,t}^{(r)}(l)$ is computed recursively from the backward as $$\beta_{i,t}^{(r)}(l)=\sum_{l_\text{next}=1}^L\beta_{i,t+1}^{(r)}(l_\text{next})\text{Pr}(\boldsymbol{x}_{i,t+1}\mid \eta_{i,t+1}=l_\text{next}, \boldsymbol{\phi})\tau_{l,l_\text{next}}^{(r)}.$$ Given the forward and backward probabilities, the classification probabilities of $\lambda_i$ is computed as $\text{Pr}(\lambda_i=r\mid \boldsymbol{x}, \boldsymbol{y}, \boldsymbol{\phi})=\frac{b_r\sum_{l=1}^L\beta_{i,T}^{(r)}(l)\alpha_{i,T}^{(r)}(l)}{\sum_{r^\prime=1}^Rb_{r^\prime}\sum_{l=1}^L\beta_{i,T}^{(r^\prime)}(l)\alpha_{i,T}^{(r^\prime)}(l)}$, and the classification probabilities of $\eta_{i,t}$ is calculated as $\text{Pr}(\eta_{i,t}=l\mid \boldsymbol{x}, \boldsymbol{y}, \boldsymbol{\phi})=\sum_{r=1}^R\frac{\beta_{i,t}^{(r)}(l)\alpha_{i,t}^{(r)}(l)}{\sum_{l^\prime=1}^L\beta_{i,t}^{(r)}(l^\prime)\alpha_{i,t}^{(r)}(l^\prime)}\text{Pr}(\lambda_i=r\mid \boldsymbol{x}, \boldsymbol{y}, \boldsymbol{\phi})$. Posterior samples involving the labels of $\lambda_i$ are then relabeled using the Stephens' relabeling algorithm based on $\text{Pr}(\lambda_i=r\mid \boldsymbol{x}, \boldsymbol{y}, \boldsymbol{\phi})$. For posterior samples involving labels of $\eta_{i,t}, t=1,\ldots,T$, relabeling is performed based on $\text{Pr}(\eta_{i,1}\mid \boldsymbol{x}, \boldsymbol{y}, \boldsymbol{\phi})$, which has been shown to be sufficient by simulation experiments. Figure \ref{fig:relabel_br} and \ref{fig:relabel_pi} demonstrates the performance of the relabeling algorithm for one simulation replication. We observe that the posterior samples exhibit good mixing after relabeling.

\section{Simulation}
\label{sec:sim}

\subsection{Data Generation}\label{sec:sim1.1}

In this section, we present the visualization of the item-response probabilities we used to generate the longitudinal item responses, the simulated individual latent class sequences, and their transitions over time, which are described in more details in Section 3 of the main text.

\begin{figure}
\centerline{\includegraphics[scale=0.6]{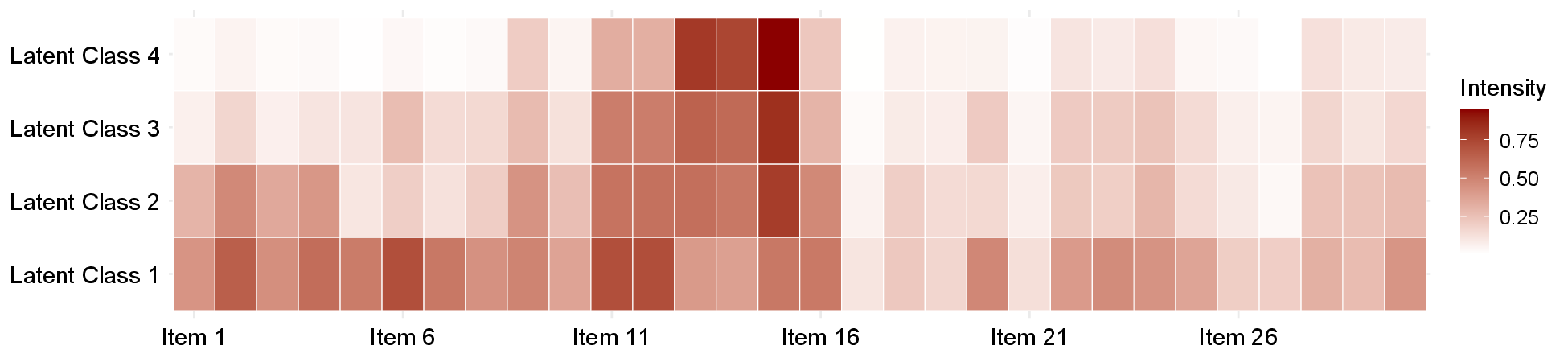}}
\caption{Heatmap of the item average intensity by the simulated item-response latent classes, with darker red representing higher intensity.} \label{fig:sim_intensity}
\end{figure}

To characterize the latent class-specific item-response probabilities $p_{lj}^{(k)}$ and to illustrate differences across latent classes, we summarize the response patterns using the average intensity of each item within each latent class. We follow a previous paper by \citet{harlow2017not} and define the average intensity of the $k$-th item in the $l$-th latent class on a scale of [0,1] as $Q_{lk}=\sum_{j=1}^{d_k}p_{lj}^{(k)}(\frac{j-1}{d_k-1})$, with values close to 1 indicating higher item (or symptom) intensity. Figure \ref{fig:sim_intensity} presents the item average intensity profiles by latent classes.

\begin{figure}
\centerline{\includegraphics[scale=0.7]{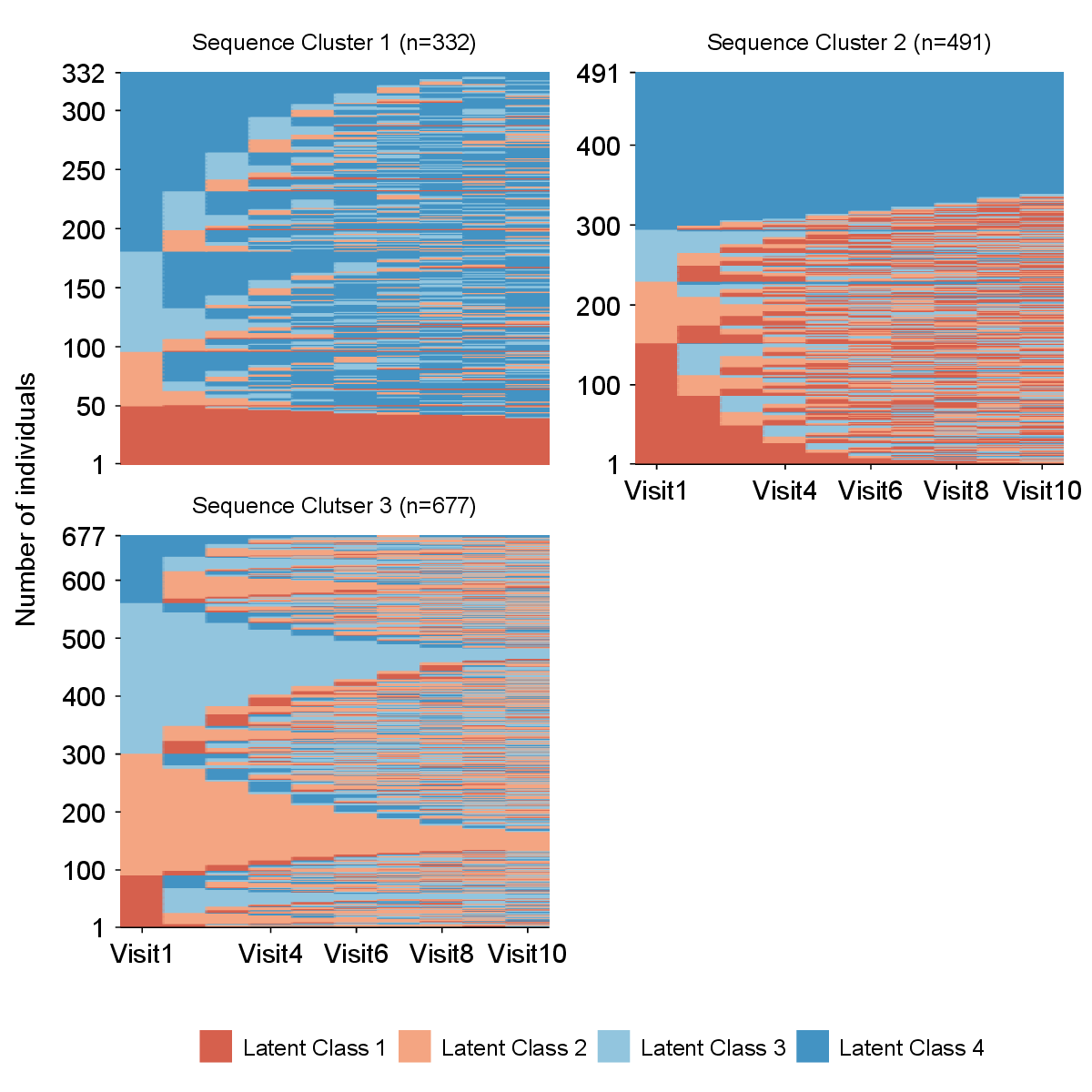}}
\caption{Visualization of latent class sequences $\{\eta_{i,1},\ldots,\eta_{i,T}\}$ by sequence clusters $\lambda_i$. Within each panel, each row depicts a latent class sequence for a simulated individual in the corresponding sequence cluster.} \label{fig:sim_sequence}
\end{figure}

\begin{figure}
\centerline{\includegraphics[scale=0.75]{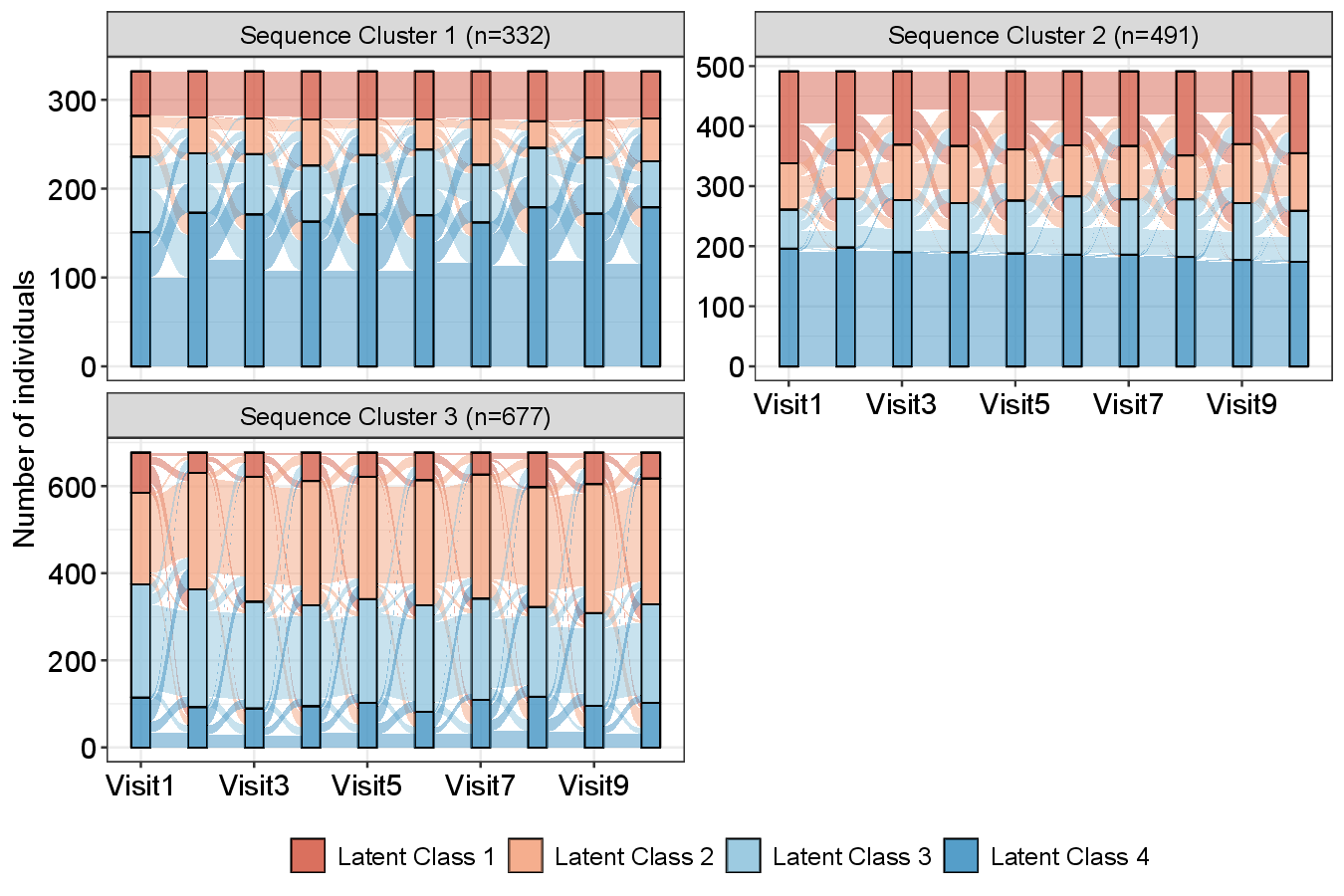}}
\caption{An alluvial diagram to visualize the latent class transitions over visits.} \label{fig:sim_alluvial}
\end{figure}

Figure \ref{fig:sim_sequence} and Figure \ref{fig:sim_alluvial} present the individual latent class sequences $\{\eta_{i,1},\ldots,\eta_{i,T}\}$ and their transitions over time, respectively. 

\subsection{Model Selection}\label{sec:sim1.2}

When demonstrating the performance of the proposed model in the simulation studies in Section 3 of the main text, we assumed known values of $R$ and $L$. In practice, however, these quantities are typically unknown and except for the prior knowledge we have about the data, we hope to employ model selection criteria to guide data-driven choices of $R$ and $L$. To this end, we evaluate and compare the model fit under different combinations of $R$ and $L$ based on their predictive accuracy. Although cross-validation is naturally the most ideal way to measure the out-of-sample predictive accuracy, the time and computational burden of our model makes it challenging. Instead, we consider several commonly used model selection criterion based on adjusted within-sample predictive accuracy with various types of bias corrections to the log pointwise predictive density (lppd) as approximations. Specifically, we examine the Akaike Information Criterion (AIC, \citet{akaike1973information}), Bayesian Information Criterion (BIC, \citet{schwarz1978estimating}), Deviance Information Criterion (DIC, \citet{spiegelhalter2002bayesian}), Widely Applicable Information Criterion (WAIC, \citet{watanabe2010asymptotic}) and approximate leave-one-out cross validation (LOO-CV) with appropriate adaptation to the Bayesian framework. 

Let $\hat{\boldsymbol{\phi}}$ denote the posterior mean of model parameters $\boldsymbol{\phi}$. AIC performs bias correction by subtracting the number of parameters estimated in the model $\delta$ from the log predictive density and is calculated as $\text{AIC} = -2\log p(\boldsymbol{x},\boldsymbol{y}\mid \hat{\boldsymbol{\phi}})+2\delta$, where under our model setting $\delta=R-1+R(L-1+L(L-1))+L\times\sum_{k=1}^K(d_k-1)+L+R-1$. BIC introduces an additional penalty depending on the sample size $N$ and is defined as $\text{BIC}=-2\log p(\boldsymbol{x},\boldsymbol{y}\mid \hat{\boldsymbol{\phi}})+\delta\log N$. DIC is a Bayesian version of AIC and adopts the same formula for bias correction. However, DIC replaces $\delta$ with a data-based effective number of parameters $p_{\text{DIC}}=2\{\log p(\boldsymbol{x},\boldsymbol{y}\mid \hat{\boldsymbol{\phi}}) - E_{\boldsymbol{\phi}}[\log p(\boldsymbol{x},\boldsymbol{y}\mid \boldsymbol{\phi})\mid \boldsymbol{x},\boldsymbol{y}]\}$, with the second term approximated using posterior samples. WAIC is a more fully Bayesian approach that corrects the effective number of parameters to adjust for overfitting and is defined as $\text{WAIC}=-2\sum_{i=1}^N\log E_{\boldsymbol{\phi}}[p(\boldsymbol{x}_i,y_i\mid \boldsymbol{\phi})\mid \boldsymbol{x},\boldsymbol{y}]+2p_{\text{WAIC}}$, where $p_{\text{WAIC}}=\sum_{i=1}^N\text{var}_{\boldsymbol{\phi}}[\log p(\boldsymbol{x}_i,y_i\mid \boldsymbol{\phi})\mid \boldsymbol{x},\boldsymbol{y}]$. The out-of-sample predictive fit from the exact LOO-CV requires refitting the model for $N$ times, but here we adopt an approximate LOO-CV calculated from the existing posterior samples using Pareto smoothed importance sampling introduced by \citet{vehtari2017practical} and the implementation is through the R package \textsf{loo} developed by \citet{gelmanloo}.

To assess the performance of these criteria, we conduct another simulation study based on the data generated in Section 3.1 of the main text with a focus on the binary outcome. We consider candidate number of sequence clusters $R_{\text{candidate}}$ from $\{2,3,4,5\}$ and candidate number of latent classes $L_{\text{candidate}}$ from $\{3,4,5\}$, with $R=3, L=4$ as the underlying true combination. At each simulation, we compute each criterion across all candidate combinations and choose the combination that minimizes the corresponding criterion value. We conduct 100 simulations in total. Table \ref{tab:sim_selection} summarizes the frequency of choosing each combination of $R$ and $L$ using the five candidate model selection criteria. AIC and BIC consistently identify the true combination under our model setting. LOO-CV tends to correctly recover $L$, with a tendency to overestimate $R$. Both DIC and WAIC favor more complex models, with high frequencies of selecting combinations with larger values of $R$ and $L$. Based on the above results, we adopt AIC and BIC to guide our model selection when it comes to the application study.

\begin{table}
\caption{Frequency of choosing the combination of $R$ and $L$ under the five model selection criterion. We omit candidate combinations that didn't got selected by any of the selection criterion. $R=3, L=4$ is the true combination. \label{tab:sim_selection}}
\renewcommand{\arraystretch}{0.9}
\begin{center}
\begin{tabular}{cccccc}
\hline
Combination of $R$ and $L$ & AIC & BIC & DIC & WAIC & LOO-CV \\
\hline
$R=3, L=4$ & 95 & 99 & 0 & 0 & 72 \\
$R=3, L=5$ & 0 & 0 & 18 & 33 & 0 \\
$R=4, L=4$ & 5 & 1 & 11 & 0 & 19 \\
$R=4, L=5$ & 0 & 0 & 25 & 39 & 0 \\
$R=5, L=4$ & 0 & 0 & 29 & 0 & 8 \\
$R=5, L=5$ & 0 & 0 & 17 & 28 & 1\\
\hline
\end{tabular}
\end{center}
\end{table}

\section{Application to the Longitudinal Symptomatology Scale Responses from SWAN}

\begin{figure}
\centerline{\includegraphics[scale=0.5]{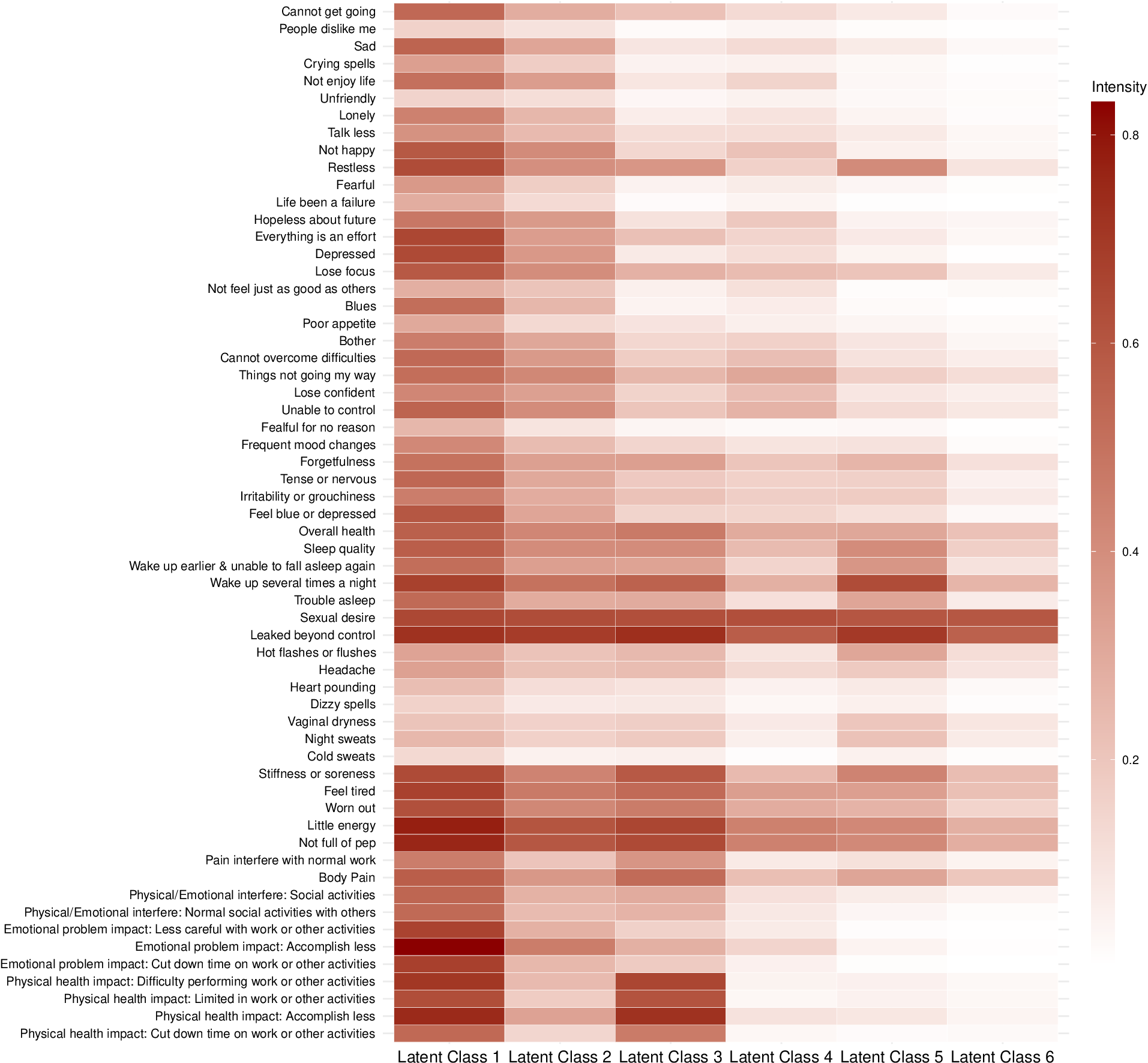}}
\caption{Heatmap of symptom intensity by the six identified item-response latent classes from the joint model with outcome of fall frequency, with darker red representing higher intensity. From left to right, the latent classes are ordered from class with most symptoms to class with least symptoms. The names on the left are shorter labels for the originally administered items.} \label{fig:swan_intensity_fallcat}
\end{figure}

\begin{figure}
\centerline{\includegraphics[scale=0.33]{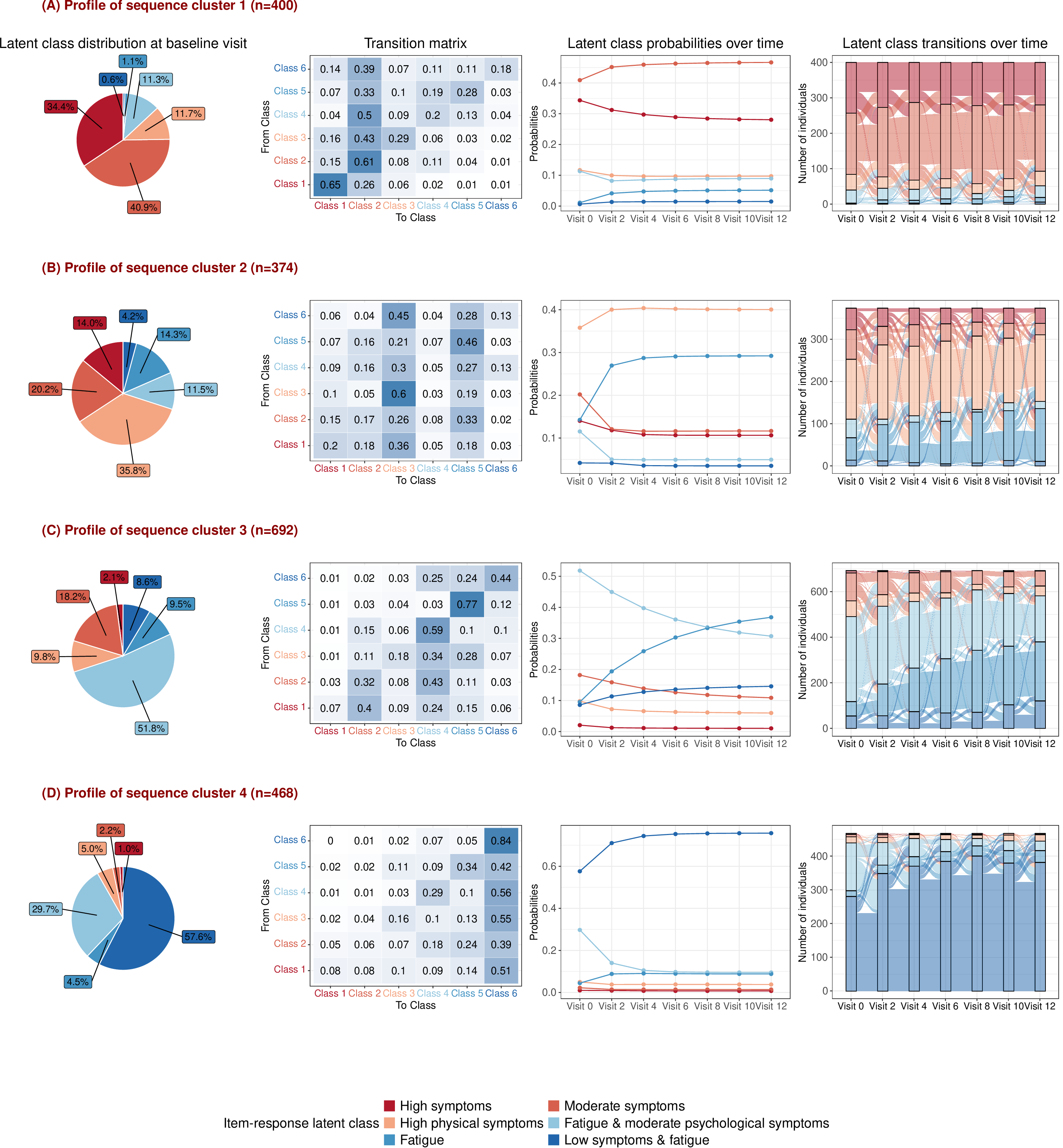}}
\captionsetup{font=footnotesize}
\caption{Profile of the four identified sequence clusters from the joint model with outcome of fall frequency. Each row presents the profile of one sequence cluster. The panels from left to right on the $r$-th row are: a pie chart representation of initial probabilities $\boldsymbol{\pi}^{(r)}$; the transition matrix $\boldsymbol{\tau}^{(r)}$, with darker blue indicating higher transition probabilities; a trajectory plot illustrating the development of latent class probabilities across visits, i.e., a representation of $\hat{\text{Pr}}(\eta_{it}=l|\lambda_i=r), l=1,\ldots,R, t = 1,\ldots,T$; and an alluvial diagram visualizing individual-level latent class transitions over time, with the individual latent class membership assigned based on modal class assignment.} \label{fig:swan_profile_fallcat}
\end{figure}

This section serves as a supplement to the analysis results for SWAN application study. Figure \ref{fig:swan_intensity_fallcat} presents the symptom intensity profiles of the six identified item-response latent classes for the outcome of fall frequency and Figure \ref{fig:swan_profile_fallcat} presents the profiles of the four identified sequence clusters.

\bibliographystyle{apalike}

\bibliography{references}